# Slow, slow, quick, quick, slow: five altmetric sources observed over a decade show evolving trends, by research age, attention source maturity and open access status

Michael Taylor, University of Wolverhampton; Digital Science

## Abstract

The study of temporal trends in altmetrics is under-developed, and this multi-year observation study addresses some of the deficits in our understanding of altmetric behaviour over time. The attention surrounding research outputs, as partially captured by altmetrics, or alternative metrics, constitutes many varied forms of data. Over the years 2008-2013, a set of 7739 papers were sampled on six occasions. Five altmetric data sources were recorded (Twitter, Mendeley, News, Blogs and Policy) and analysed for temporal trends, with particular attention being paid to their Open Access status and discipline. Twitter attention both starts and ends quickly. Mendeley readers accumulate quickly, and continue to grow over the following years. News and blog attention is quick to start, although news attention persists over a longer timeframe. Citations in policy documents are slow to start, and are observed to be growing over a decade after publication. Over time, growth in Twitter activity is confirmed, alongside an apparent decline in blogging attention. Mendeley usage is observed to grow, but shows signs of recent decline. Policy attention is identified as the slowest form of impact studied by altmetrics, and one that strongly favours the Humanities and Social Sciences. The Open Access Altmetrics Advantage is seen to emerge and evolve over time, with each attention source showing different trends. The existence of late-emergent attention in all attention sources is confirmed.

## Acknowledgements

Particular thanks are due to the late Professor Henk Moed, who, on a frosty day in Wolverhampton, prompted me to investigate altmetrics. Further thanks are due to Professor Michael Thelwall for his kind help and guidance throughout my PhD studies; to Kathy Christian at Digital Science, and Brad Allen, Ron Daniels and Lisa Colledge at Elsevier for facilitating this research, and to the reviewers for their helpful suggestions. Data was accessed from

Altmetric.com (http://www.altmetric.com); Mendeley (http://www.mendeley.com); Crossref (http://http://www.crossref.org) and Dimensions (http://www.dimensions.ai).

## Keywords

Scientometrics, altmetrics, Twitter, Mendeley, social impact, longitudinal study, policy, open access, open access altmetrics advantage, grey literature

## Introduction

The term 'altmetrics' was introduced in 2010 in the Altmetric Manifesto (Priem, Taraborelli, Groth, & Neylon, 2010) to bring together the study of web-based attention to research under one term, to "reflect the broad, rapid impact of scholarship". Hitherto, research in this area had been conducted under the umbrella term 'webometrics' (Almind & Ingwersen, 1997), which had evolved from the fields of bibliometrics and scientometrics. The introduction of the new term extended this field of study, which had largely been focused on the analysis of web hyperlinks, web citations and usage (Bar-Ilan 2000; Thelwall, 2000) to include social media and other similar data sources. This reflected the increasingly important role that social media was playing in the dissemination of research (Sugimoto, Work, Larivière, Haustein 2017).

The Altmetrics Manifesto described a number of possible advantages of altmetrics over citations. One anticipated benefit was the relative speed at which altmetrics could indicate a paper's importance, compared to traditional citation-based measures. Whereas the process of publishing and citation can be slow – and resistant to change (Wheatley & Grzynszpan, 2002) – altmetrics could be used to understand impact at, or shortly after, a paper's publication.

Despite this key focus in the Altmetrics Manifesto, temporal issues have not received significant attention from researchers. Ortega (2018) offered an analysis of six metrics over 24 months, three of which (Blogs, Mendeley and Twitter) are usually counted as altmetrics. Mendeley readership starts to accrue shortly after publishing, and to persist over a long timeframe (Maflahi and Thelwall, 2018). Temporal metrics have been calculated for twelve altmetric data sources for a twelve month period (Fang and Costas, 2020). These and other relevant studies are reviewed below.

### Field and attention source dependencies

Altmetric sources are recognized to be largely heterogenous (Haustein, 2016) and to show significant discipline differences (Zahedi, Costas and Wouters, 2014).

Past research has adequately described differences in overall usage: the two most populous data sources are Mendeley and Twitter (Thelwall, Haustein, Larivière, & Sugimoto, 2013), and these have correspondingly been the focus of most altmetric studies.

A set of papers published between 2012-18, sampled in 2019 showed high degrees of variability between fields, per altmetrics attention source (Fang et al., 2020). Over 40% of Social Sciences & Humanities and Biomedical & Health Sciences were active on Twitter; compared with 36% Life and Earth Sciences; 22% Physical Sciences and Engineering and 11% Mathematics and Computer Sciences. Mendeley showed a coverage of close to 90% for all fields. News, Blogs and Policy sources showed a similar pattern, with the Social Sciences and Humanities having a coverage of 5%, 6% and 3% respectively; Biomedical & Health Sciences, 6%, 4%, 2%; Life and Earth Sciences, 4%, 6%, 1%; Physical Sciences and Engineering 2% for News and Blogs and >0.5% for Policy; and Mathematics and Computer Sciences being below 0.5% for all three indicators.

Earlier research from 2014 sheds some light on earlier trends (Zahedi, Costas and Wouters, 2014): Mendeley showed high levels of attention in the Medical and Life Sciences, and Natural Sciences, with 50 per cent and 32 per cent of active documents coming from these areas, Arts and Humanities was represented by less than 1 per cent each; Twitter showed a similar focus, with 42 and 49 percent, and 2% respectively. Despite their relative importance (Phillips, Kanter, Bednarczyk, & Tastard, 1991; Williams, 2018), News, Blogs and Policy sources have not been subject to as much research, although a bias towards Life Sciences and Medicine has been previously demonstrated (Shema, Bar-Ilan, Thelwall 2012).

The degree to which altmetric data is reported and available to analysis is known to vary over time. Suppliers, such as Altmetric LLP, may add or remove data sources to their services. Indicators such as Wikipedia and Patent citations are relatively new to Altmetric (2015 and 2018 respectively). Furthermore, improvements to the collection and parsing processes are frequently made, which may result in increased coverage and accuracy: for example, Altmetric's news parser was improved in 2014, and Wikipedia coverage was enhanced in 2020-21, by the addition of previously unindexed languages[1]. In contrast, LinkedIn and Weibo have both removed access to their data (2013, 2015 respectively), while Google+ was discontinued in 2019[2]. Furthermore, the introduction of the General Data Protection Regulation in 2018 (European Union, 2016) requires platforms such as Twitter to remove content and data (such as Tweets and Twitter Account information) from all systems, and obliges providers such as Altmetric LLP to follow suit, with the consequence that numbers may reduce.

[1] Private communication between the author and colleagues at Altmetric LLP
[2] https://www.altmetric.com/about-our-data/our-sources/

## Open Access and the Open Access Altmetrics Advantage

Since 2001, Open Access (OA) publishing has gained momentum (Suber 2012), with sustained growth rates reported over the last decade. The proportion of OA publications was calculated to be 20% in 2009 (Björk et al., 2010), and 45% in 2015 (Piwowar, Priem, Larivière, Alperin, Matthias, Norlander et al. 2018). The scholarly database Dimensions reports that annual proportion of OA publications exceed 50 per cent of all articles and preprints in 2018[3]. The COVID-19 crisis of 2020-21 is seen as having further driven the adoption, acceptability and internationality of OA publishing (Lee and Haupt, 2020). Recent investigations have focused on the so-called Open Access Altmetrics Advantage (OAAA), finding OA articles (Holmberg, Hedman, Bowman, Didegah & Laakso 2020) and OA books (Taylor, 2020) tend to receive higher rates of social media and public engagement than non-OA research, Nevertheless, there are both disciplines and attention sources with an Open Access Altmetrics *Disadvantage* (OAAD), for example Psychology research articles linked on Twitter (Holmberg, Hedman, Bowman, Didegah & Laakso 2020) or Humanities books cited in Wikipedia (Taylor, 2020).

## The relationships between citation and altmetric data

Initial research focused on understanding the correlations between citations and various altmetrics; however the nature of the data sources, temporal issues and degree of coverage give the results an imprecise interpretation (Thelwall, 2016). As the field has developed, new techniques have been employed to understand the complex effects that social media is having on research communications (Ebrahimy, Mehrad, Setareh, & Hosseinchari, 2016).

Both citation and altmetric data are heavily skewed, with a large proportion of published outputs having low rates of activity, and a very small number tending to attract very high rates of activity.

It has been recognized that citation-based metrics vary significantly over time for discipline (Thelwall & Sud, 2016). Metrics suppliers have created article-based citation metrics that take published year and discipline into account, by computing values that present ratios of observed citations to expected citations, for publications of a particular type, publication year and subject area (Hutchins et al., 2016; Zanotto & Carvalho, 2021). Similar normalization techniques have been proposed for altmetrics (Thelwall, 2017). Positive evidence for longitudinal effects and systemic trends in different altmetric sources would be additional evidence to support the need for a similar, systematic approach to analysing altmetrics, normalized for time and discipline.

[3] https://app.dimensions.ai/analytics/publication/open_access_status/timeline?or_facet_publication_type=article&or_facet_publication_type=proceeding&or_facet_publication_type=preprint

Research into citation patterns and download patterns have demonstrated a quantifiable and interactive relationship (Moed, 2005), with both downloads acting as a leading indicator to citations (Watson, 2009), and citations leading to increased downloads (Schlögl, Gorraiz, Gumpenberger, Jack, & Kraker, 2014). Thus, we understand that the relationship between citations and downloads is dynamic and complex, even many years after publication. The complexity of the communication system is underlined by the observations that Twitter activity increases traffic to journal websites (Hawkins et al., 2014), and that increased page views can lead to increased citations (Perneger, 2004).

Scientometric researchers have observed that there are outliers, in terms of citation performance over time (Braun, Glänzel, & Schubert, 2010), and the connection between late citation emergence and altmetric data has been explored (Hou et al., 2020). Previously known as "sleeping beauties", these papers have been observed to lie dormant for a period of time (when their cohort are usually active), before becoming more highly cited, against the trend for their discipline and publication year. Identifying papers with delayed citation emergence has been proposed as a method of discovering 'hidden' or latent research (Demaine, 2018). Similar phenomena and utility may be hypothesized in altmetric data, where papers become active against cohort trend. For example, Twitter is usually seen as an early source of attention: a paper that receives tweets several years after publication might merit such a definition and be considered worthy of additional attention.

## Known longitudinal observations for altmetrics

The original Manifesto listed a number of potential sources of altmetric data other than Mendeley: in particular, Twitter, and blogs. Although we consider all of these different data sources under the common term of 'altmetrics', there are many differences between them: including audience, purpose, methodology and access. These contribute towards the differences between how quickly these data appear, how quickly their activities peak, and the degree to which their activity is sustained over time.

Tweets are one of the quickest indicators to appear (Ortega, 2018), accumulating within a few days of an article becoming available. High rates of Twitter activity have been shown to correlate well with later rates of citation, for a small group of medical articles, (Eysenbach, 2011). Twitter activity is usually considered to be relatively short-lived, both for preprints (Shuai, Pepe, & Bollen, 2012) and published papers.

Mendeley readership is known to correlate well with citation rates (Thelwall, 2018) and academic usage (Mohammadi, Thelwall and Kousha, 2016), and is closely related to usage and downloads (Kudlow et al., 2017). Mendeley readership counts start to accumulate very soon after publishing, and even before articles are officially published (Maflahi and Thelwall, 2018) but persist over a prolonged timeframe (Maflahi and Thelwall, 2018), having a strong relationship over time with both citations and downloads (Ortega, 2018).

In a retrospective analysis covering altmetrics in the first twelve months after publication, social media and blog attention was observed to appear soon after publication before dropping away, whereas Mendeley readership continued to accrue (Ortega, 2018). Blogging activity continues many years after publication, (Jamali & Alimohammadi, 2015), and drives page views and downloads of the original article (Allen, Stanton, Di Pietro, & Moseley, 2013). News and Twitter coverage have been shown to be related to the newsworthy qualities of the article, rather than as a result of interaction between the two attention sources (Htoo et al., 2022). Similarly, a relationship between COVID-19 articles receiving news and blog attention has been observed (Fraumann & Colavizza, 2022).

Direct research into how altmetrics vary over time and by age is relatively underdeveloped. A variety of temporal metrics were calculated for twelve altmetric indicators over the course of a year, characterising Twitter, News and Blogs as *fast* attention sources, whereas Policy and Wikipedia were characterised as being *slow* (Fang and Costas, 2019). Although there has been research that reported a temporal advantage for *usage* data (Zhang, Wang, Xie, Du et a, 2020), there has been no research into the temporal nature of the OAAA.

## Objectives

This paper addresses a number of gaps in the literature by examining data collected over a multi-year period for five different altmetric indicators: Mendeley, Twitter, News, Blogs and Policy documents. The analysis explores some of the trends exposed in the data, providing insights into the development of both behaviour and platform usage, as well as longitudinal variation by attention source, Open Access status and discipline.

1. Current knowledge about altmetric trends over time is partial, with most research studying data within a short period after publication. This research addresses this gap by studying trends over a multi-year timespan.
2. Although it is understood that platform usage and collection techniques show temporal trends, this has not been examined in an academic context over a multi-year timescale.
3. Existence of late-emergent research has been observed using citation analysis, but the altmetric equivalent has only been hypothesized: this research investigates the possible existence of this phenomena.
4. The existence of an OAAA has been previously confirmed, however the degree to which it developed over the last decade has not previously been investigated. This research explores the dynamics of the OAAA over the last decade.

# Method

## Data

This research was initiated during the Snowball Metrics project (Clements et al., 2017), and used a set of 7739 primary research articles drawn from a larger set being used in the development of research publication metrics (Taylor, 2014). All were authored by researchers affiliated with UK institutions participating in the project, and selected as representative of the outputs of participating institutions (University of Oxford, University College London, University of Cambridge, Imperial College London, University of Bristol, University of Leeds, Queen’s University Belfast, and University of St Andrews). In general, research articles with authors affiliated with these universities have very high rates of altmetric activity: in 2018, the proportion of these papers with activity reported by Altmetric was approximately twice the global average.

The selected papers had publishing dates ranging from 2008 to 2013, and therefore show a range of ages across the duration of the observation period (2013-2021): a paper published in 2011 would be 7 years old at the time of the 2018 sample. The set of papers published in 2013 is smaller than other years, as this year was incomplete at the time of the observation.

Data was retrieved from the Altmetric and Mendeley APIs using research licenses, on six occasions from 2013 to 2021, (September 2013, September 2014, April 2017, June 2018, September 2020 and June 2021).

Altmetric collects data from a variety of sources, including Twitter (by detecting links to research in tweets); news (links and parsed mentions of research from over 2000 news sources); blogs (links and parsed mentions from a list of approved research-focussed blogs) and policy papers (links and parsed citations from a list of policy repositories). Mendeley allows users to save links to research outputs and makes totals of these per document available via an API.

Subject area metadata was retrieved for each article from Digital Science's Dimensions API. Dimensions assigns articles into Fields of Research classifications by a machine learning process at a paper rather than a journal level, thus allowing for greater granularity of analysis. The process assigns up to four subject codes per paper using title and abstract text, where available (Herzog, Sorensen, & Taylor, 2016). To increase the sample sizes for greater statistical power, the Fields of Research codes were further grouped into four larger disciplines – Physical and Technological Sciences (PTS), Life Sciences (LS), Medical and Health Sciences (MHS) and Humanities and Social Sciences (HSS).

The Dimensions method does not always assign a subject code to a research publication: either the machine learning system doesn't not produce a high enough certainty to meet the threshold, or insufficient text is available to the classification process. Of the 7739 papers, 14.5% did not have a Field of Research code in Dimensions and were classified following the predominant classification of their journal. The distributions for both published year and discipline are presented in Table 1.

***Table 1** Number of papers in each cohort, by date published, discipline and Open Access status*

| Discipline | Published year | | | | | | Total |
|---|---|---|---|---|---|---|---|
| | 2008 | 2009 | 2010 | 2011 | 2012 | 2013 | |
| Physical and Technological Sciences (PTS) | 121 | 157 | 221 | 534 | 586 | 97 | 1716 |
| Life Sciences (LS) | 204 | 244 | 324 | 565 | 487 | 143 | 1967 |
| Medical and Health Sciences (MHS) | 302 | 359 | 451 | 915 | 769 | 199 | 2995 |
| Humanities and Social Sciences (HSS) | 105 | 127 | 161 | 321 | 263 | 84 | 1061 |
| **Total** | **732** | **887** | **1157** | **2335** | **2105** | **523** | **7739** |
| Open Access | 454 | 582 | 797 | 1541 | 1325 | 343 | 5042 |
| Closed | 278 | 305 | 360 | 794 | 780 | 180 | 2697 |

Data was also broken down by Open Access status, as reported by Unpaywall in 2021. The period that papers were selected from had relatively low rates of OA publishing, however, papers often become OA over time. The observation period took place at a time when OA publishing rates were growing strongly (Appendix, Figure 12). Papers were classified as *either* OA (being either Gold, Green or Bronze) *or* Closed.

## Analysis

Three calculations were used to compare the relative performance of the cohorts:

1. The percentage of papers with any reported altmetric activity.
2. Average values for the two high-frequency attention sources, Mendeley readers and unique Twitter accounts.
3. To compare OA and non-OA papers, percentage coverage and average values were calculated for OA and non-OA papers (as above), with the value for the OA cohort being divided by the non-OA cohort to calculate an OAAA (Taylor, 2020).

The preferred way of calculating an average for non-normally distributed data is to use a geometric mean. Altmetric data, in common with citation data, is highly skewed, with a small number of papers typically getting a disproportionate amount of attention. The approach taken here uses a log mean approach to counter the skewness of these two attention sources (Thelwall & Fairclough, 2015). News, policy and blog citations typically occur at very low frequencies, rendering comparison of average values of limited use.

To test for significance, the proportion of papers with altmetric activity for each age were tested using a Chi-squared test, comparing the actual observed proportion of papers with attention against the overall average for that cohort. To test the number of Mendeley readers and Twitter accounts, mean values were calculated from the natural log, and evaluated using an ANOVA 1-way test. To test for the significance of the OAAA, we evaluated the OA and non-OA cohorts for significance, using t-tests for the Mendeley and Twitter means and z-tests for the proportions of populations with attention.

# Results

## Longitudinal trends over the observation period

### Mendeley

The proportion of articles with at least one Mendeley reader for the dataset was high in the year of publication (at least 77.8%) and increased over time to be consistently above 99% after 5 years (Table 2). For the most recent set of papers, published in 2013, Mendeley reached 97.5% coverage within one year of publication. For the three paper ages containing four cohorts, there was insufficient evidence of a difference between cohorts in the proportion with at least one Mendeley reader.

***Table 2** Proportion of papers with Mendeley readers over the observation period (%)*

| Pub year | Age of paper at observation in years | | | | | | | | | | | | | |
|---|---|---|---|---|---|---|---|---|---|---|---|---|---|---|
| | 0 | 1* | 2* | 3* | 4* | 5* | 6- | 7- | 8- | 9- | 10- | 11- | 12- | 13 |
| 2008 | | | | | | 85.1 | 98.9 | | | 99.3 | 100.0 | | 100.0 | 100.0 |
| 2009 | | | | | 84.0 | 99.1 | | | 99.8 | 100.0 | | 100.0 | 100.0 | |
| 2010 | | | | 87.6 | 98.7 | | | 99.0 | 99.9 | | 100.0 | 100.0 | | |
| 2011 | | | 90.2 | 98.7 | | | 99.5 | 99.7 | | 100.0 | 100.0 | | | |
| 2012 | | 87.0 | 98.0 | | | 99.3 | 99.4 | | 99.4 | 100.0 | | | | |
| 2013 | 77.8 | 97.5 | | | 99.6 | 99.6 | | 99.8 | 99.8 | | | | | |

** Years in which the proportion of papers with Mendeley activity differed between cohorts for articles of the same age as calculated by a chi-squared test, p >= 0.05.*
*- Insufficient evidence to reject the null hypothesis*

Sustained growth in average readers across the lifetime is shown for each of the cohorts of papers (Table 3). The rate at which papers acquire additional readers on Mendeley appears to decline towards the end of the observation period. In the year between the first two observations (12 months, between 2013–14), the approximate annual growth varies from 6.8 (papers published in 2012) to 19.2 (2009); between the second two observations (15 months, between 2017–18), from 8.5 to 14.6, reported growth for the final observation (9 months between 2020–2021) is lower, from 3.4 to 6.0.

***Table 3** Average values based on ln(1+x) Mendeley readers over the observation period*

| Pub year | Age of paper at observation in years | | | | | | | | | | | | | |
|---|---|---|---|---|---|---|---|---|---|---|---|---|---|---|
| | 0 | 1* | 2* | 3* | 4* | 5* | 6* | 7* | 8* | 9* | 10* | 11* | 12- | 13 |
| 2008 | | | | | | 16.4 | 33.0 | | | 59.0 | 73.6 | | 116.9 | 122.1 |
| 2009 | | | | | 14.9 | 34.1 | | | 62.6 | 75.7 | | 122.3 | 128.2 | |
| 2010 | | | | 16.6 | 32.5 | | | 59.4 | 73.9 | | 120.0 | 125.7 | | |
| 2011 | | | 11.9 | 20.6 | | | 42.1 | 51.5 | | 79.8 | 83.2 | | | |
| 2012 | | 7.9 | 14.7 | | | 35.1 | 42.3 | | 68.0 | 71.5 | | | | |
| 2013 | 4.5 | 13.3 | | | 39.1 | 47.6 | | 75.6 | 80.1 | | | | | |

** Years in which the average number of Mendeley readers differed between cohorts for articles of the same age as calculated by a one-way ANOVA, p >= 0.05.*
*- Insufficient evidence to reject null hypothesis*

By reading down the columns, we can compare a set of like-for-like documents at similar post-publication ages. Papers published in 2008, five years after publication have an average of 16.4 readers; papers published in 2013 have an average of 47.6 readers after five years: Mendeley usage almost tripled between 2013- 2018. In contrast, the growth in year 9 (representing the time period 2018-2021) is lower, approximately 1.2.

Mendeley coverage is consistently close to 100% for all disciplines from the second observation onwards.

Average Mendeley readership varies strongly by discipline (Figure 1), with the Life Sciences (LS) having the most readers: approximately twice the number of readers than the lowest, Physical and Technical Sciences (PTS). Medical and Health Sciences (MHS) and the Humanities and Social Sciences (HSS) show similar values. The apparent reduced rate of reader acquisition between 2020–2021 is represented across all disciplines. Nevertheless, all disciplines acquire Mendeley readers across the entire observation period. In general, the older papers have higher rates of readership: the biggest difference being shown in the HSS discipline. PTS show the smallest variation between publication years.

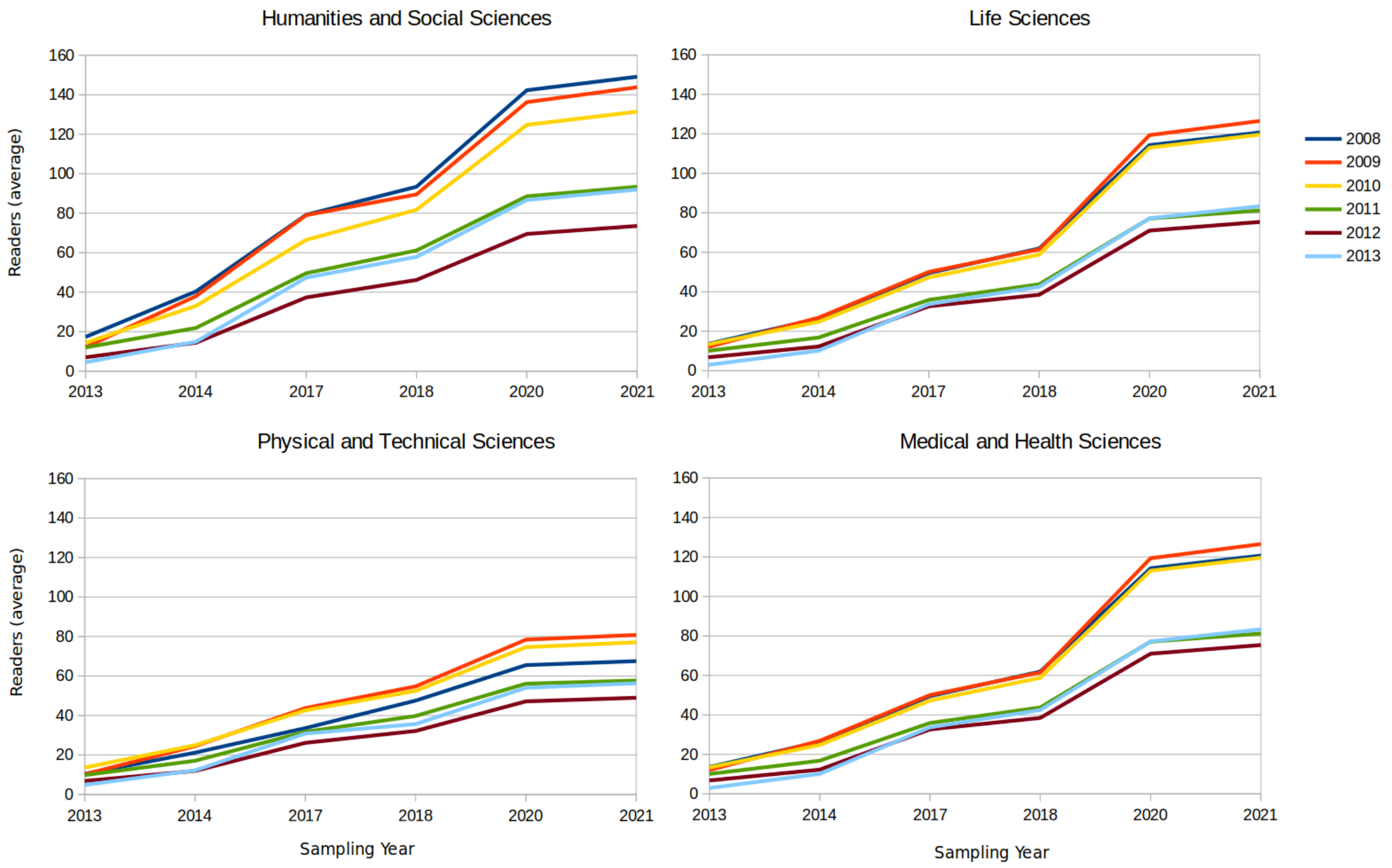

***Fig. 1*** *Average Mendeley readers per paper over the observation period by discipline for the six cohorts of papers*

## Twitter

The proportion of articles with at least one Tweet varies by both age of publication, and the relative maturity of the Twitter platform (Table 4). After five years, 31% papers published in 2008 had received attention on Twitter. In contrast, over 90% of papers published in 2013 had received attention on Twitter the year after publication. As Twitter coverage of newer papers increased, so did the coverage of older papers, with coverage of 2008 publications rising from 31% in 2013 to 50%  in 2021, at which stage the publications were 13 years old. The older sets of papers are more likely to receive late first Tweets than younger.

***Table 4*** *Proportion of papers with Tweets over the observation period (%)*

| Pub year | Age of paper at observation in years | | | | | | | | | | | | | |
|---|---|---|---|---|---|---|---|---|---|---|---|---|---|---|
| | 0 | 1- | 2* | 3* | 4 | 5* | 6* | 7* | 8* | 9* | 10* | 11- | 12- | 13 |
| 2008 | | | | | | 31.1 | 38.7 | | | 45.5 | 46.4 | | 49.7 | 50.1 |
| 2009 | | | | | 35.1 | 43.3 | | | 48.4 | 50.7 | | 55.9 | 56.1 | |
| 2010 | | | | 41.0 | 48.9 | | | 55.2 | 58.0 | | 60.2 | 60.9 | | |
| 2011 | | | 70.8 | 77.2 | | | 78.4 | 78.1 | | 77.4 | 77.4 | | | |
| 2012 | | 84.5 | 92.4 | | | 92.5 | 90.8 | | 90.8 | 89.0 | | | | |
| 2013 | 76.5 | 90.6 | | | 90.8 | 90.6 | | 90.2 | 90.2 | | | | | |

** Years in which the proportion of papers with Twitter activity differed between cohorts for articles of the same age as calculated by a chi-squared test, p >= 0.05.*
*- Insufficient evidence to reject the null hypothesis*

The growth rate of both Twitter coverage and Twitter averages decreases over time, with coverage for the oldest set of papers stabilising at ~50%, with an average of 1 unique Twitter account per paper. In contrast, the newest papers stabilise at ~90% with a mean of ~3.5 accounts (Table 5). This observation is supported by the coverage data: for papers published in 2012 (where 2013 is the first full year), coverage is approaching 100% by 2014.

The rate of Twitter growth is observed to decrease: for papers aged five years old, coverage grew 2.9 times (31.1%-90.6%) between 2013-2018; average Twitter accounts sharing links to the same papers, at the same time, grew 9 times (0.4-3.6).

***Table 5*** *Average values based on ln(1+x) Twitter accounts per paper over the observation period (%)*

| Pub year | Age of paper at observation in years | | | | | | | | | | | | | |
|---|---|---|---|---|---|---|---|---|---|---|---|---|---|---|
| | 0 | 1* | 2* | 3* | 4* | 5* | 6* | 7* | 8* | 9* | 10* | 11* | 12* | 13 |
| 2008 | | | | | | 0.4 | 0.5 | | | 0.8 | 0.9 | | 1.0 | 1.1 |
| 2009 | | | | | 0.4 | 0.6 | | | 0.9 | 1.1 | | 1.3 | 1.3 | |
| 2010 | | | | 0.5 | 0.8 | | | 1.1 | 1.3 | | 1.4 | 1.5 | | |
| 2011 | | | 1.3 | 1.7 | | | 1.9 | 1.9 | | 1.9 | 1.9 | | | |
| 2012 | | 2.0 | 2.6 | | | 2.7 | 2.7 | | 2.7 | 2.5 | | | | |
| 2013 | 2.1 | 3.4 | | | 3.6 | 3.6 | | 3.4 | 3.4 | | | | | |

** Years in which the average number of Twitter accounts differed between cohorts for articles of the same age as calculated by a one-way ANOVA, p >= 0.05.*
*- Insufficient evidence to reject null hypothesis*

Over the observation period, coverage of the four disciplines grows at different rates, suggesting differences in Twitter sharing (Figure 2). For all four disciplines, the younger papers have both the highest coverage, and highest average Twitter accounts (Figure 3). PTS has the highest difference in coverage, but the lowest variation of averages by publication year. All four disciplines show retrospective first Twitter attention for the older cohorts, although this is most marked for HSS and LS papers. With the exception of HSS papers published in 2012[4] , the average accounts tweeting about research is remarkably consistent, being generally fewer than three accounts per paper. Although the coverage is seen to increase, this average does not, suggesting a commensurate growth in the number of active accounts, as well as the number of papers tweeted about.

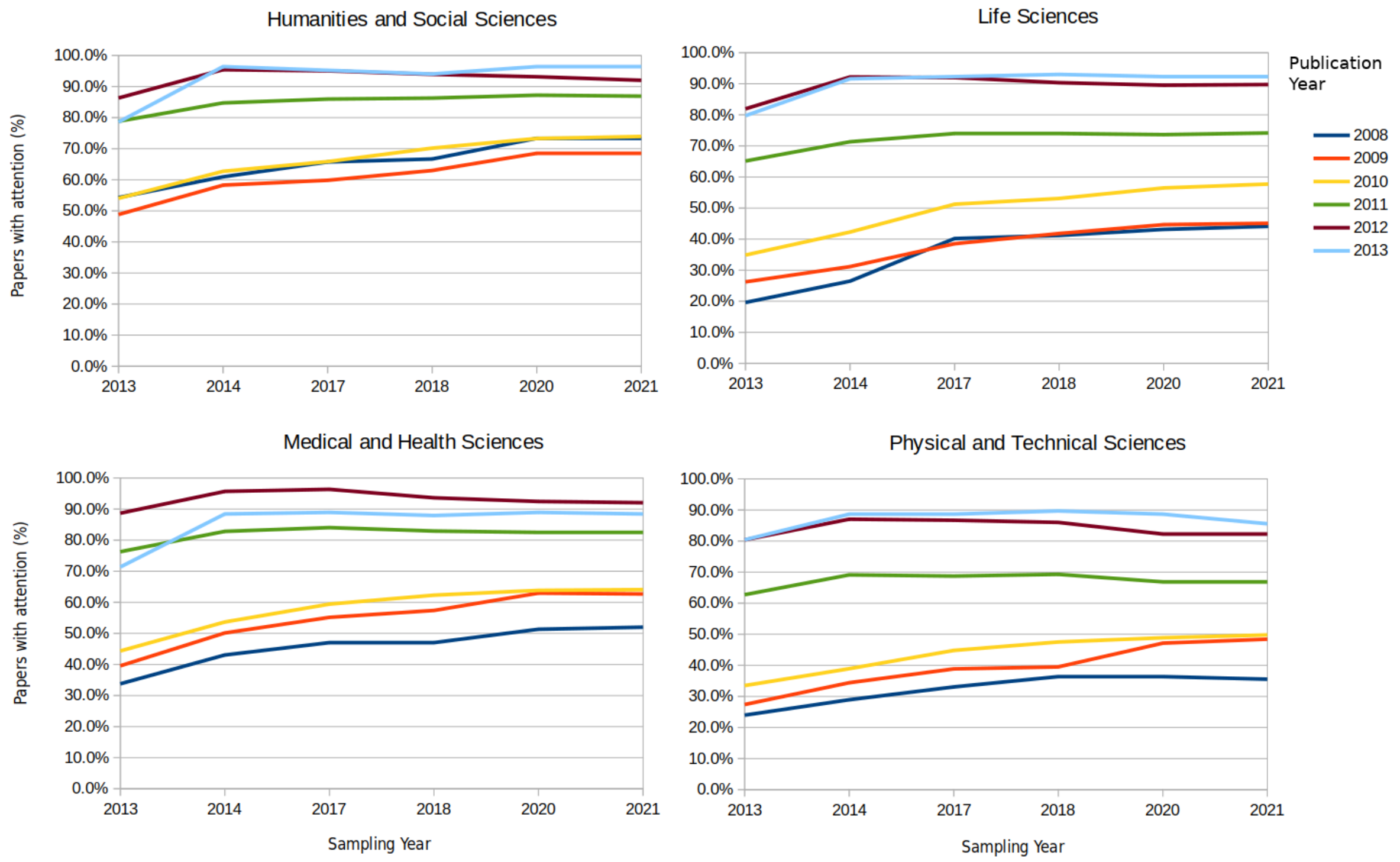

***Fig. 2** Proportion of papers with Tweets over the observation period by discipline (%)*

[4] The Twitter average for Humanities and Social Sciences papers published in 2012 is skewed by four highly tweeted papers, viz, https://doi.org/10.1007/s00213-012-2657-5; https://doi.org/10.1098/rspb.2011.1373; https://doi.org/10.1371/journal.pone.0031824; https://doi.org/10.1371/journal.pmed.1001244; together they have ~1000 Tweets

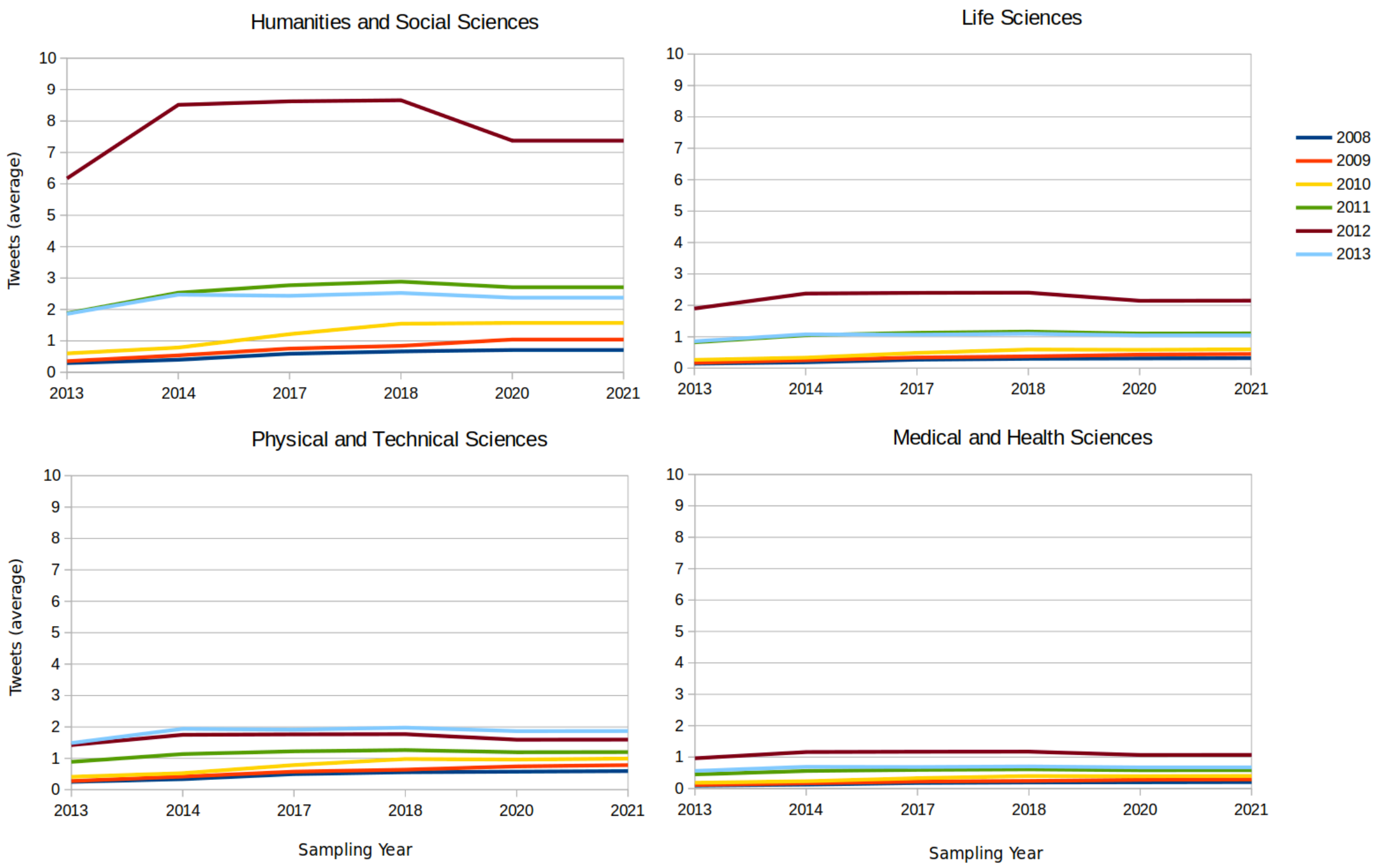

***Fig. 3*** *Average unique Twitter accounts over the observation period by discipline*

## News

The proportion of articles with at least one news mention is seen to grow over time, before levelling off towards the end of the observation period (Table 6). As with other attention sources, the proportion of papers being covered by news rises more significantly for the older papers than the newer: coverage for 2008 publication grows approximately 5 times as the articles age from 5-13 years; whereas 2013 publication only doubles through years 0-8. Nevertheless, the coverage for all publications falls within a reasonably narrow range: from 15.3% to 23.3%.

Altmetric improved their news collection process in 2014[5] (years 5 and 6 for the 2008 cohort, 4 and 5 for the 2009 cohort etc), which may explain the dramatically higher rates between the first two samples. Comparing news coverage for papers at ages 8 (14.0-20.7%) and 9 (14.3-18.7%), - all of which fall after the improvement - suggests that news coverage has not, generally, increased independently of the change.

***Table 6*** *Proportion of papers with news attention over the observation period (%)*

| Pub year | Age of paper at observation in years |
|---|---|

[5] Altmetric added an NLP process to their mention parsing process that doubled precision. Private correspondence.

| | 0 | 1* | 2* | 3* | 4* | 5* | 6* | 7* | 8* | 9- | 10* | 11- | 12- | 13 |
|---|---|---|---|---|---|---|---|---|---|---|---|---|---|---|
| 2008 | | | | | | 4.1 | 7.5 | | | 14.3 | 16.9 | | 19.9 | 20.6 |
| 2009 | | | | | 5.5 | 9.4 | | | 16.2 | 18.7 | | 22.4 | 23.3 | |
| 2010 | | | | 6.2 | 10.5 | | | 16.2 | 18.1 | | 21.1 | 21.7 | | |
| 2011 | | | 4.4 | 8.2 | | | 11.9 | 13.2 | | 15.3 | 15.3 | | | |
| 2012 | | 4.6 | 8.9 | | | 11.9 | 14.0 | | 14.0 | 15.6 | | | | |
| 2013 | 10.3 | 15.3 | | | 17.8 | 18.5 | | 20.7 | 20.7 | | | | | |

** Years in which the proportion of papers with news attention differed between cohorts for articles of the same age as calculated by a chi-squared test, p >= 0.05.*
*- Insufficient evidence to reject the null hypothesis*

News coverage for all four disciplines grows over time, with HSS and MHS articles continuing to gain first coverage at a steady rate (Figure 4). Although PTS and LS receive most news coverage in 2013, their rate of growth is less over the observation period, allowing MHS, and HSS to have overtaken them by the 2020 observation.

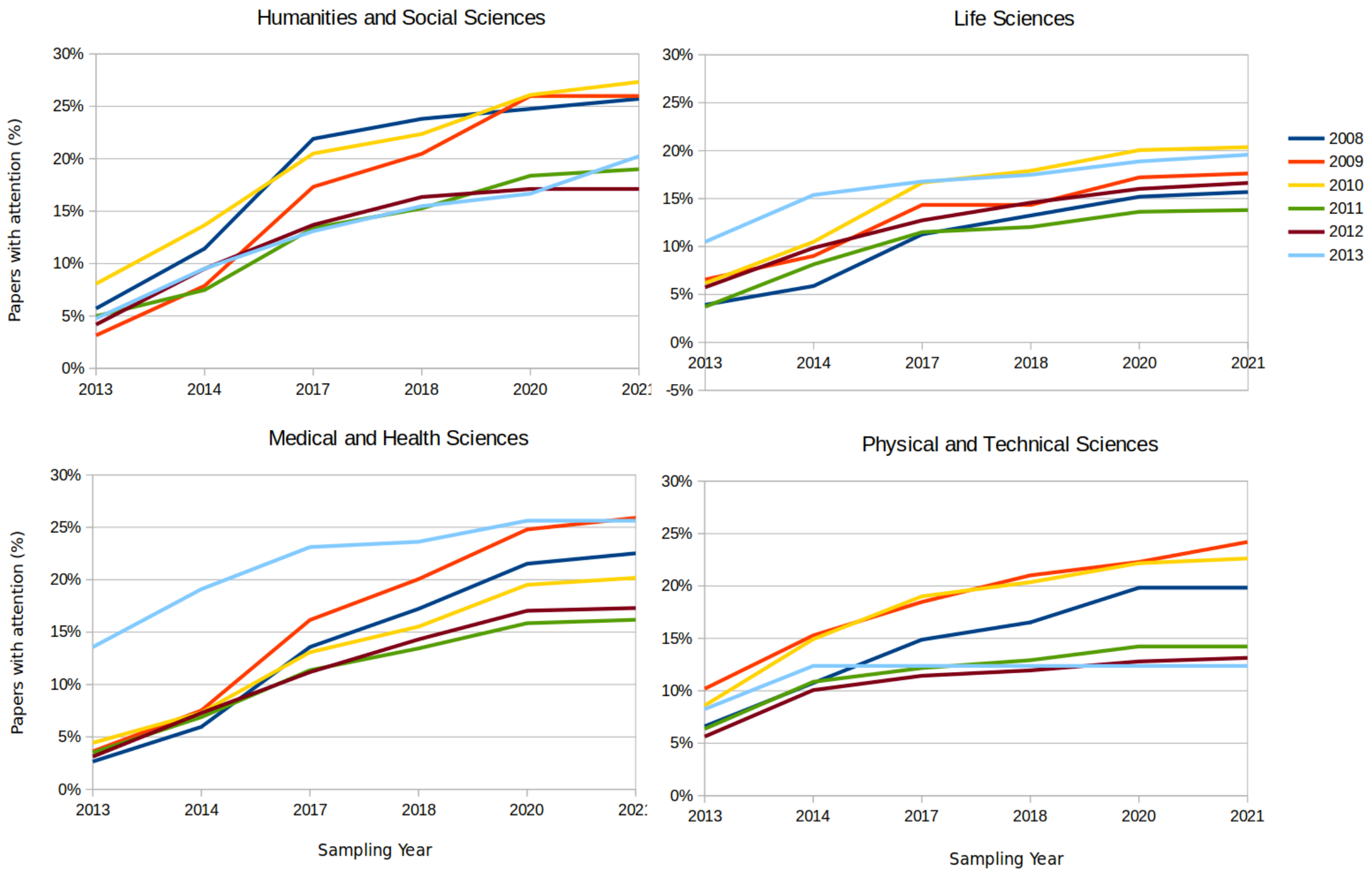

***Fig. 4*** *Proportion of papers with news attention over the observation period by discipline (%)*

## Blogs

The proportion of articles with at least one blog mention consistently grew as they aged (Table 7). The oldest articles grew from 38-49.7% between years 5-13; the youngest grew from 8.8-21.2% between years 0-8. In contrast with other indicators, blogging coverage decreased over the observation period, falling from 38% for the 2008 cohort, in year 5, to just over 20% for the corresponding aged papers published in 2013. At age 5, coverage of the oldest papers is approximately twice that of the newest, at ages 8 and 9, the difference is greater than twice.

***Table 7*** *Proportion of papers with blog coverage over the observation period (%)*

| Pub year | Age of paper at observation in years | | | | | | | | | | | | |
|---|---|---|---|---|---|---|---|---|---|---|---|---|---|
| | 0 | 1* | 2* | 3* | 4* | 5* | 6* | 7* | 8* | 9* | 10* | 11- | 12- | 13 |
| 2008 | | | | | | 38.0 | 42.8 | | | 45.8 | 47.4 | | 49.3 | 49.7 |
| 2009 | | | | | 36.4 | 42.7 | | | 46.1 | 48.6 | | 50.1 | 50.2 | |
| 2010 | | | | 32.8 | 38.8 | | | 41.8 | 44.1 | | 45.1 | 45.3 | | |
| 2011 | | | 20.7 | 24.5 | | | 26.9 | 28.0 | | 28.9 | 28.9 | | | |
| 2012 | | 11.0 | 15.2 | | | 17.5 | 19.3 | | 19.3 | 20.7 | | | | |
| 2013 | 8.8 | 14.3 | | | 17.6 | 20.5 | | 21.2 | 21.2 | | | | | |

** Years in which the proportion of papers with blog coverage differed between cohorts for articles of the same age as calculated by a chi-squared test, p >= 0.05.*
*- Insufficient evidence to reject the null hypothesis*

Blogging coverage is shows significant disciplinary differences: HSS grow disproportionately faster, from being one of the lowest blogged disciplines, to one of the highest by the end of the observation period (Figure 5). The other fields grow more slowly, with the MHS being the least covered, suggesting a lower likelihood that research will get a late first mention on blogs.

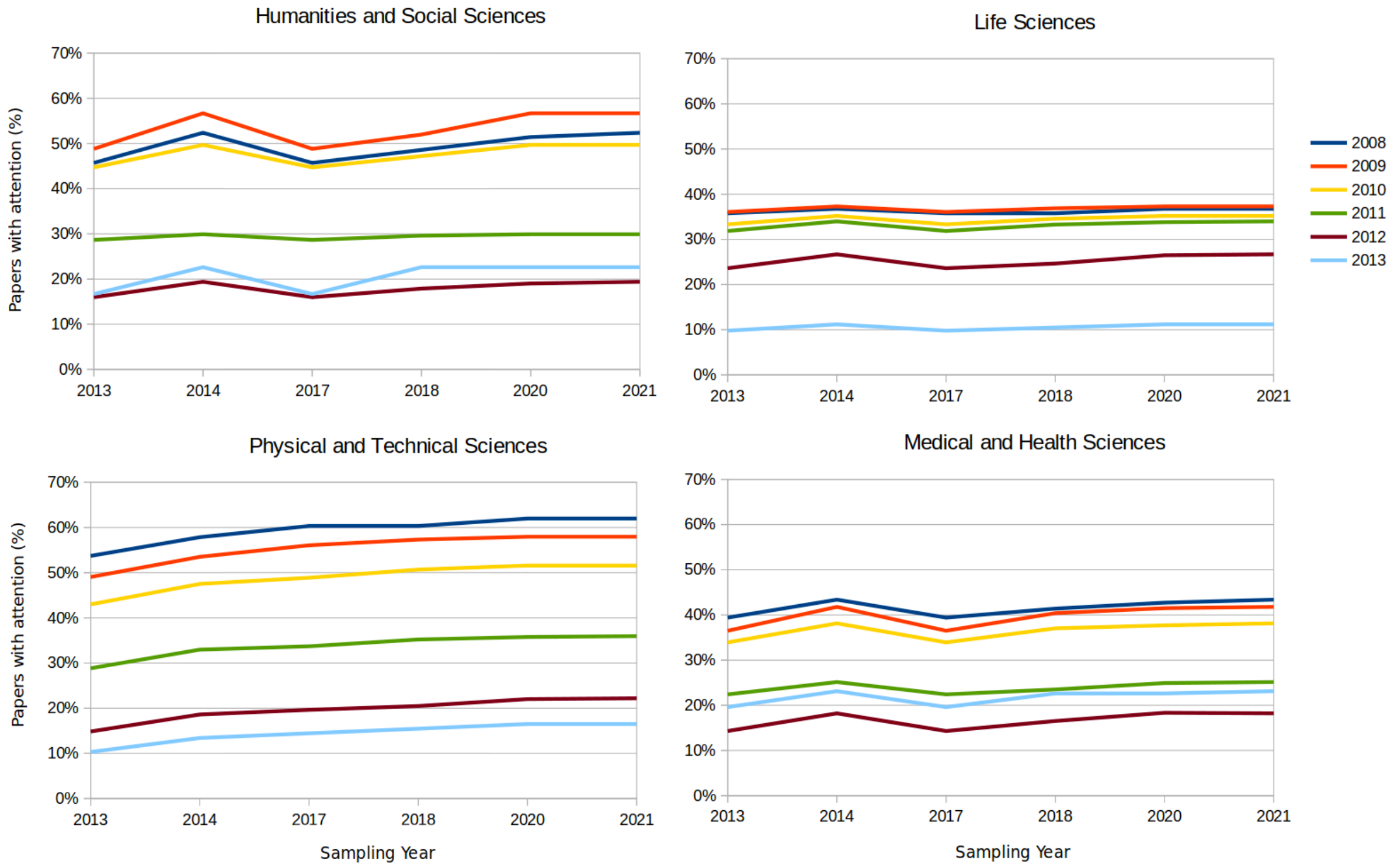

***Fig. 5*** *Proportion of papers with blog coverage over the observation period by discipline (%)*

## Policy

The proportion of articles with Policy citations is the smallest reported in this research with no or few citations appearing in the first few years after publication. However, it shows the most marked growth as publications age, typically reaching 10% coverage towards the end of the first decade. (Table 8). Policy attention is the only data source that shows a growth in probability of novel attention as a research paper ages, with peak probability of receiving a first citation being between years 5 and 7 after publication. Although the overall probability is very low, papers previously uncited by policy documents continue to receive new citations in the thirteenth year after publication.

***Table 8*** *Proportion of papers with policy citations over the observation period (%)*

| Pub year | Age of paper at observation in years | | | | | | | | | | | | | |
|---|---|---|---|---|---|---|---|---|---|---|---|---|---|---|
| | 0 | 1- | 2- | 3- | 4- | 5* | 6* | 7- | 8* | 9* | 10* | 11- | 12- | 13 |
| 2008 | | | | | | 1.1 | 2.2 | | | 11.6 | 14.8 | | 17.8 | 18.6 |
| 2009 | | | | | 1.5 | 3.6 | | | 12.9 | 14.0 | | 18.3 | 19.3 | |
| 2010 | | | | 1.5 | 2.4 | | | 8.9 | 10.9 | | 16.2 | 17.5 | | |
| 2011 | | | 0.5 | 1.0 | | | 5.5 | 7.8 | | 11.5 | 11.5 | | | |
| 2012 | | 0.1 | 0.7 | | | 4.4 | 6.3 | | 6.3 | 9.3 | | | | |
| 2013 | 0.0 | 0.0 | | | 3.1 | 5.2 | | 8.8 | 8.8 | | | | | |

** Years in which the proportion of papers with policy citations differed between cohorts for articles of the same age as calculated by a chi-squared test, p >= 0.05.*
*- Insufficient evidence to reject the null hypothesis*

Policy coverage shows the most marked longitudinal and discipline variation, with almost zero attention shown to research until reaching its fifth year (Figure 6). MHS and HSS dominate Policy coverage showing coverage at or around 20% by 2021. PS and LS appear to stabilize relatively quickly at a much lower rate.

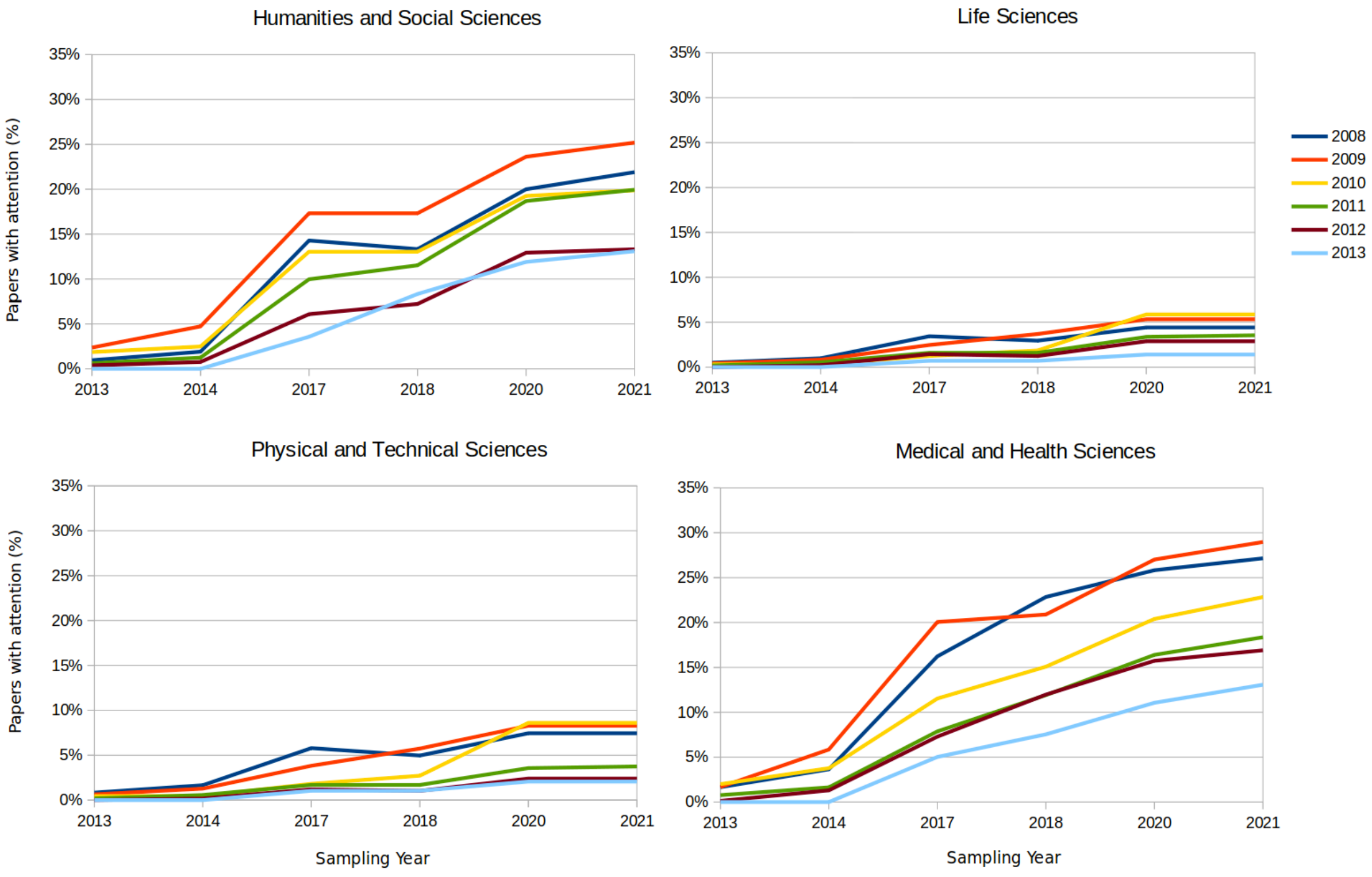

***Fig. 6** Proportion of papers with policy coverage over the observation period by discipline (%)*

## Open Access Altmetrics Advantage

As Mendeley coverage is almost complete for all cohorts, the ratio of coverage from OA- to non-OA papers approximates to 1.00 for all samples. However, a consistent OAAA for the mean number of Mendeley readers exists, ranging from 1.17 to 1.47, suggesting that OA papers have been saved more frequently than their non-OA counterparts (Figure 7).

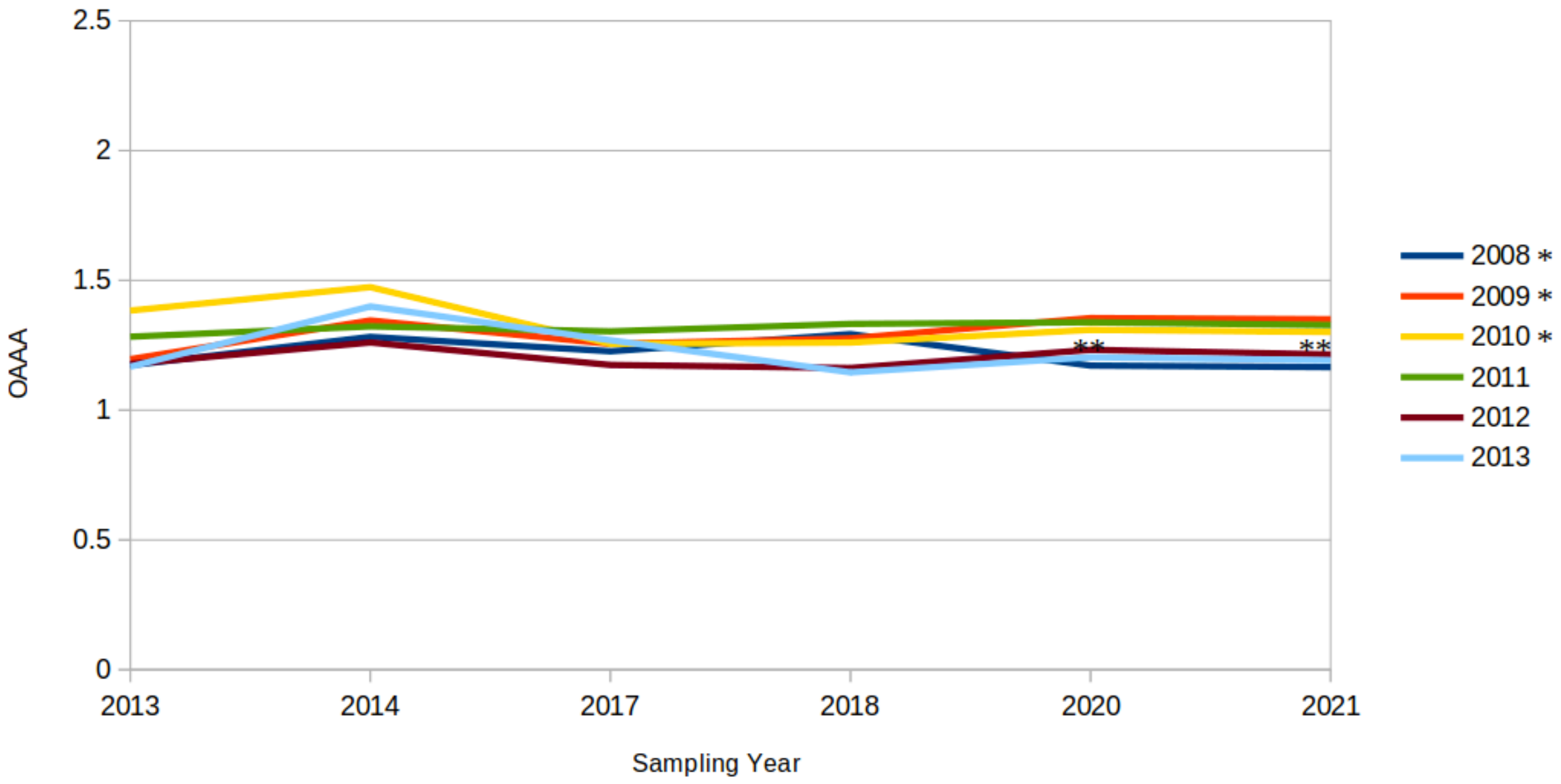

***Fig. 7*** *Ratio average Mendeley readers, OA publications: non-OA publications (OAAA)*
** Samples (and whole years) are significant as calculated by t-test, p >= 0.05.*

For Twitter coverage, older OA papers appear to have an OAAA of less than one (i.e. are less likely to be tweeted about), however this disadvantage is seen to decrease and ultimately disappear over the observation period. In general, the values for mean Twitter attention are more variable than either Twitter coverage, or Mendeley averages: the older papers (published in 2008-2010) show a tendency to move from either parity or having a disadvantage early in the experiment, to having a moderate advantage by 2021 (Figure 8).

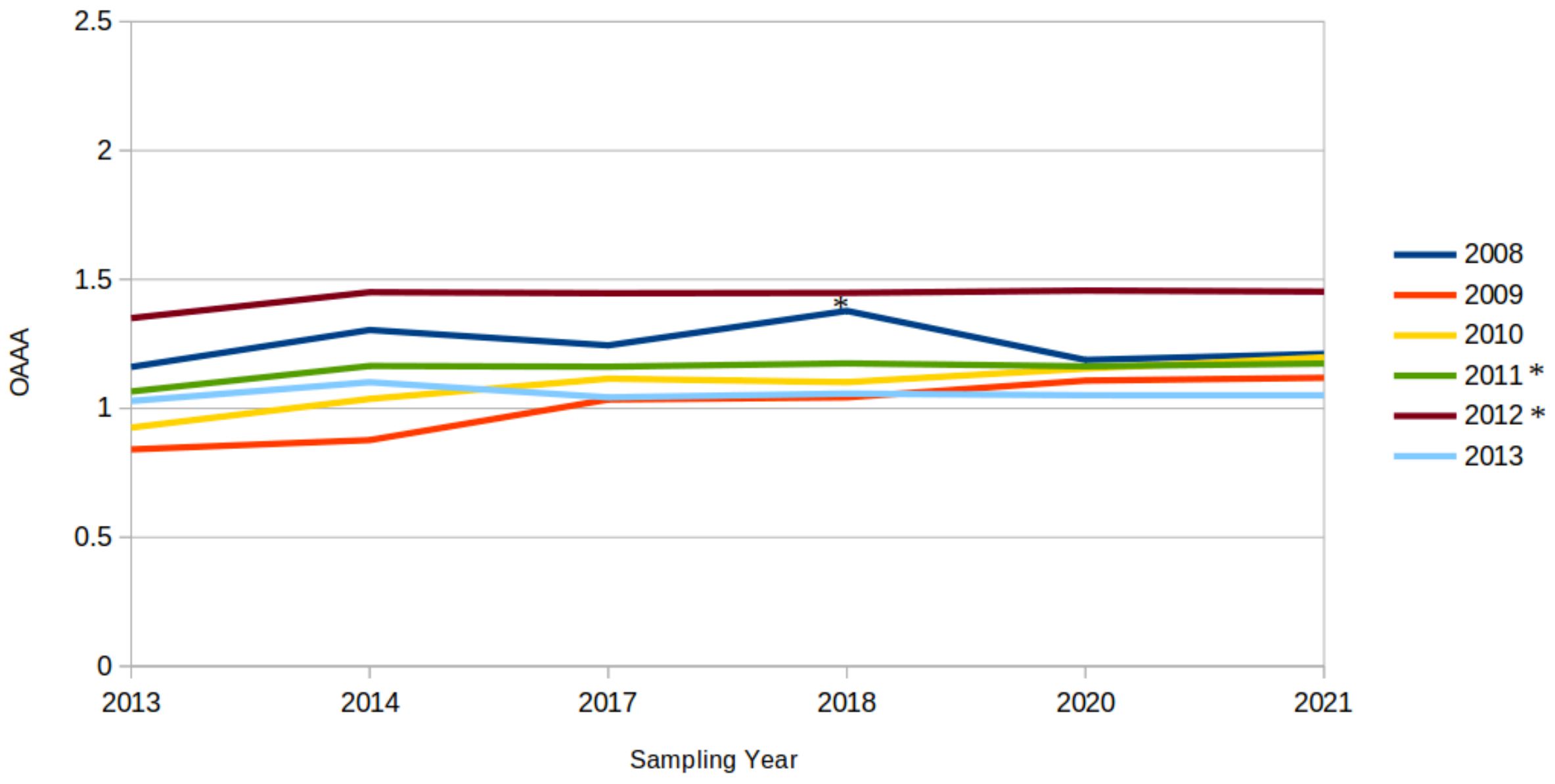

***Fig. 8*** *Ratio of average Twitter accounts for OA publications: non-OA publications (OAAA)*
** Samples (and whole years) are significant as calculated by t-test, p >= 0.05.*

This trend, that the OAAA varies by both sampling period and publication date is reflected in attention from News outlets, with OA papers published between 2008-2010 showing a striking disadvantage at the start of the experiment but showing an advantage at the end (Figure 9). Papers published in 2011-2013 both start and finish with an OAAA, whereas the older papers start with a disadvantage, before gaining their advantage.

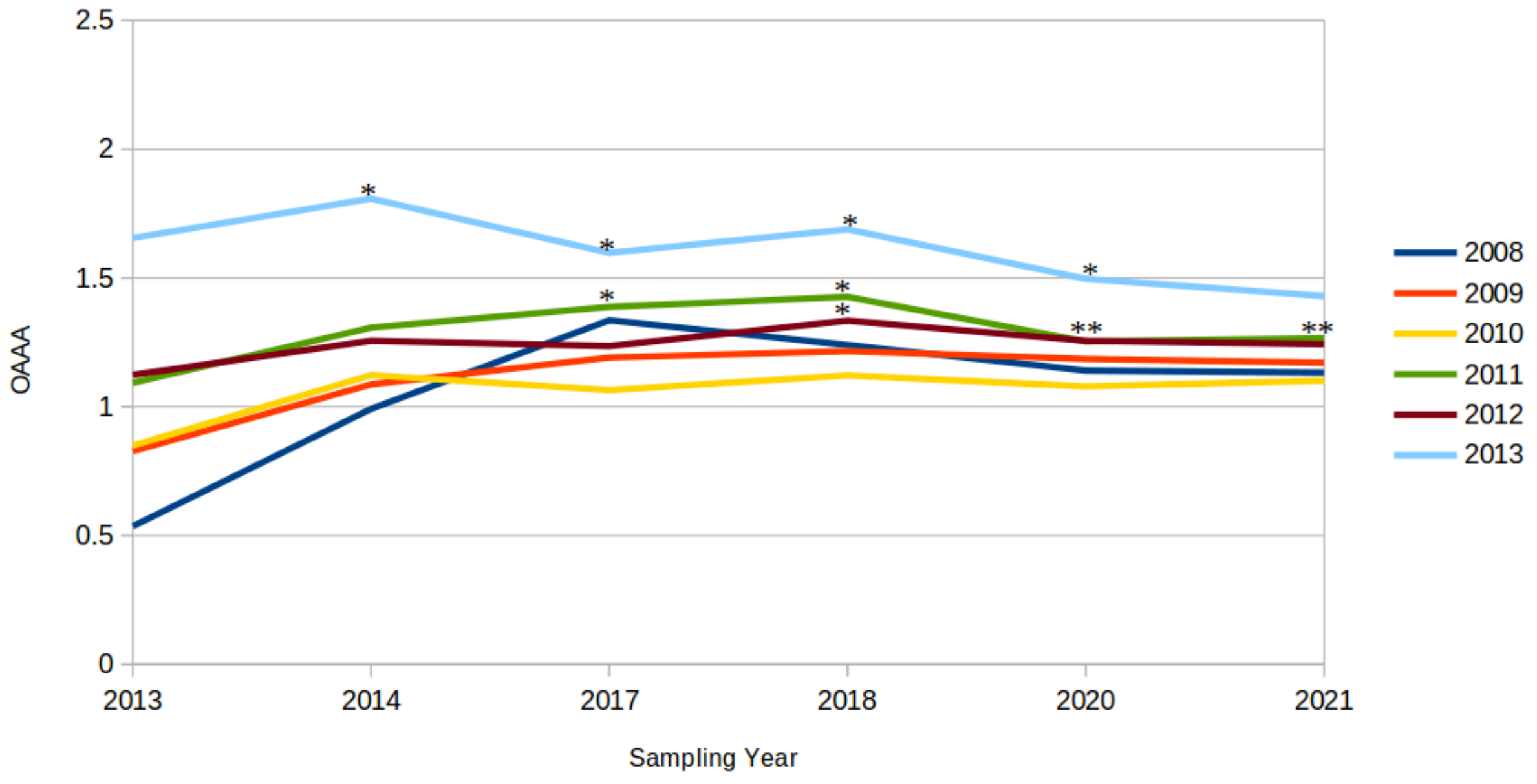

***Fig. 9*** *Ratio of news coverage for OA publications : non-OA publications (OAAA)*
** Samples (and whole years) are significant as calculated by z-test, p >= 0.05*

Blog coverage, in contrast, shows an OAAA for all data points, albeit it one that is stronger for the younger papers (Figure 10). In striking contrast to other data points, the OAAA is generally seen to decrease over the sampling period for the younger papers, published in 2012-2013. The OAAA is generally consistent across the sampling period for the older papers.

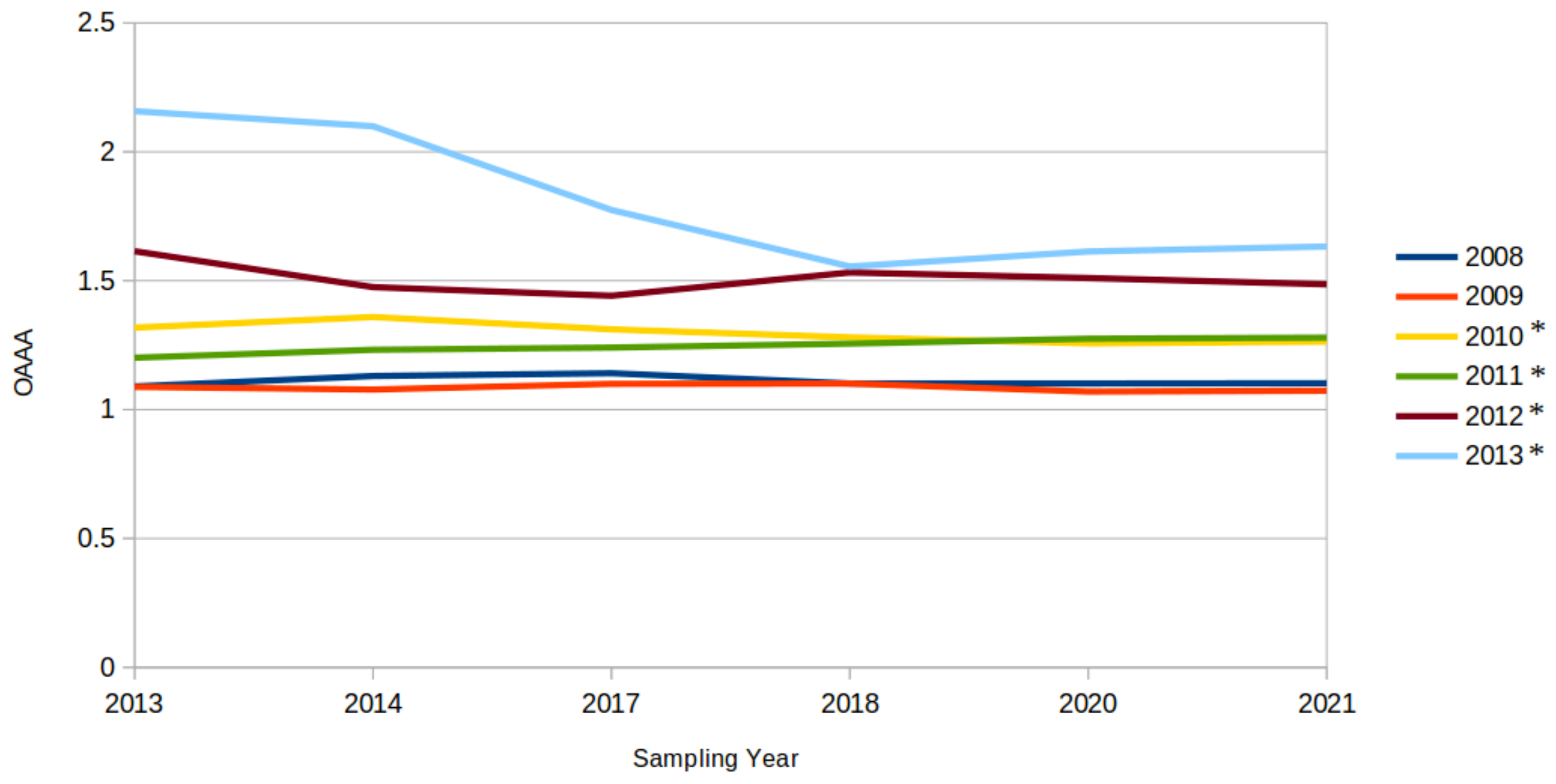

***Fig. 10*** *Ratio of blog coverage for OA publications : non-OA publications (OAAA)*
** Samples (and whole years) are significant as calculated by z-test, p >= 0.05*

Policy coverage shows a marked different in trend from other attention sources, all papers of all ages having an OAAA of less than 1, i.e. a disadvantage (Figure 11). This disadvantage is seen to shrink over the course of the experiment, nevertheless, most cohorts show an OAAA disadvantage even in the final year, with only papers published in 2009 showing an advantage, and 2008 and 2011 showing neither advantage new disadvantage.

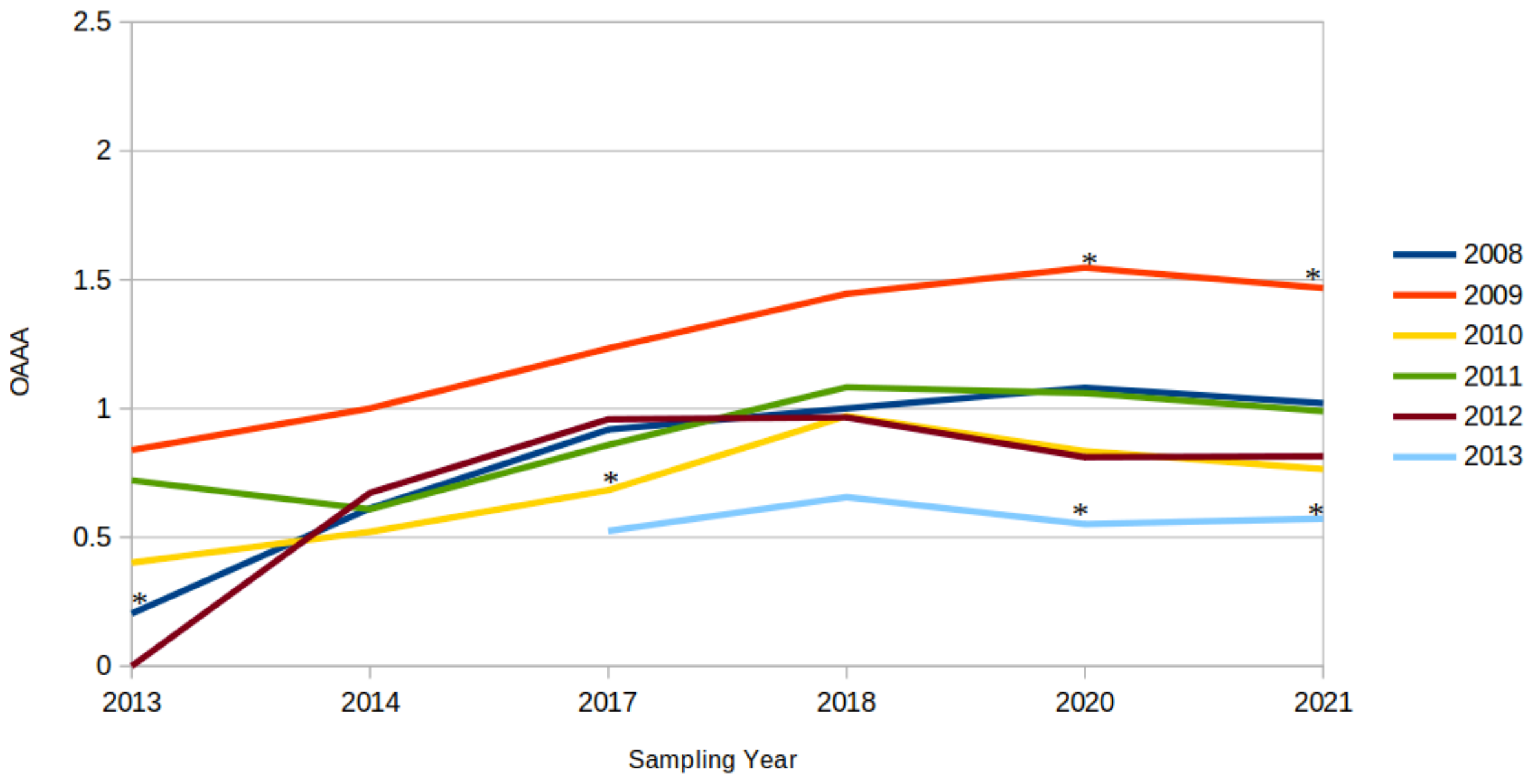

***Fig. 11*** *Ratio of policy coverage for OA publications : non-OA publications (OAAA)*
** Samples (and whole years) are significant as calculated by z-test, p >= 0.05.*

## Discussion

There are a number of limitations in this paper. First, the samples are unequally distributed across the age, which limits the effectiveness of statistical analysis. Secondly, the OA status of the articles is that defined by Unpaywall in 2021. It is likely that a subset of these papers have become open over time, however there is no reliable data on which to identify those dates. This analysis compares non-OA research with all OA, without analysing for differences between the different forms of OA publication.

The researchers' affiliations for this dataset are drawn from institutions that are known to show very high rates of altmetrics activity, and therefore care should be taken when extrapolating findings to other research outputs.

Only five altmetric data sources are examined. Although, for example, Altmetric have added Wikipedia data retrospectively, it wasn't available at the start of the observation period. Whilst this paper does contain some insights from Altmetric LLP on changes to the data collection process, it cannot report on smaller, day-to-day changes that may result in changes in data collection trends.

### Variations between altmetrics suppliers and attention sources

Variations exist between the methods used by different altmetric providers to collect and collate data, and between different data sources collected by individual altmetric data providers.

In terms of data collection and analysis, some data sources have Application Programme Interfaces (APIs) that enable direct querying of a database to obtain data using a key, such as a Digital Object Identifier. Other sources are significantly more complex and need to be 'scraped', i.e, the documents accessed from the web, and then parsed for mentions and citations. In the case of Mendeley, where the catalogue of documents is, to some extent, crowd-sourced, the data is aggregated against an individual document key by an automated process and accessed via an API. In the case of Twitter, News and Policy documents, the burden for aggregating data is placed with the supplier, e.g. PlumX or Altmetric. Data aggregators may choose to report only DOIs or URLs embedded in the documents, or they may develop complex technologies to identify, extract and match text-only mentions.

These processes all have strengths and weaknesses: use of a data source from an API typically doesn't expose the underlying data. Web content that has to be 'scraped' is subject to a number of external factors that affects both the number of documents with attention, and the rate at which that attention is reported. These variables may be commercial, legal or technical. Unlike scholarly content, blogs and newspapers are rarely archived by the publishers and may become unavailable over time. Political and organizational changes often result in unstable web repositories of policy documents and research data (Eng, 2017), and unclear copyright and poor hosting and archive practices reduce the likelihood of policy documents remaining available for ongoing examination or analysis.

The process of, for example, identifying a citation in a policy document – where there are no formal standards for making citations - is heuristic. Suppliers such as Altmetric invest in algorithms to identify and resolve these citations; and will engage in ongoing projects to improve both precision and recall figures. These improvements may lead to fluctuations in altmetric activity being reported, even in the absence of an underlying causative trend. These improvements may be retrospectively applied to improve historical data.

## Growth in coverage over time

The sampling process gives us the opportunity of comparing sets of papers at the same stage of maturity in several years. Three ages have four samples. Year 5 is represented by the 2008, 2009, 2012 and 2013 cohorts, being sampled in 2013, 2014, 2017 and 2018 respectively. Year 8 is represented by 2009, 2010, 2012 and 2013, and Year 9 by the 2008, 2009, 2012 and 2013 cohorts. Only one cohort was sampled in the year of publishing (2013), and one thirteen years after publishing (2008's cohort, sampled in 2021). Using these data points, we can analyse the growth (or otherwise) of each attention source.

Mendeley coverage approaches saturation for all except for the very youngest papers in the earliest sampling periods. Nevertheless, the growth in the average Mendeley readership is significant, showing sustained year-on-year growth in the years following the platform's launch. Growth is particularly noticeable in samples from 2014, following Elsevier's 2013 acquisition of Mendeley, and the relatively short-lived #mendelete campaign (Deville, 2018). An unknown question is whether Mendeley users have remained loyal to the platform as their careers have developed, and whether Elsevier has continued to grow its market share: the average readership growth between 2020 and 2021 appears to be at a much lower rate than may have been expected, for example, the 2013 cohort were 7 and 8 years old in 2020 and 2021, and grew by an average of only 4.6 readers. In contrast, the 2010 cohort were 7 and 8 in 2017 and 2018, and grew by an average of 14.5. This apparent reduction in readership acquisition in latter years is seen across all four disciplines, and reinforces previous observations of a potential decline in Mendeley usage (Fang et al., 2020).

Twitter's growth in research coverage is dramatic: the 2008 cohort achieved 31.1% coverage in its fifth year; the 2013 cohort is recorded with 90.6% coverage at the same age. The mean Tweets per paper also increase, the 2008 cohort having 0.4 tweets per paper in its fifth year; the 2013 having 3.6. Average tweets per paper are approximately twice as high for MHS and HSS than for the other disciplines. Usage of the Twitter platform has expanded significantly since its launch, and this growth continued throughout the sampling period, with figures of 7.8% in 2015 and 5.4% in 2019 (Statista, 2021). These findings are inline with other research (Fang et al, 2020), although offer new insights in how adoption of this platform expanded rapidly, before slowing its rate of growth over the observation period.

News coverage growth reported by Altmetric.com is likely to result from improvements in their collection processes and through the increasingly availability of news on the Internet. News coverage of research favours younger publications: papers aged 5, sampled in 2013, 2014, 2020 and 2021 jump from 4.1% to 18.5%. In contrast, papers aged 9, sampled in 2013, 2014, 2017 and 2018 show no significant growth.

Blog coverage is observed to decline over time: all measurements show a year-on-year decline. The oldest papers aged 9, sampled in 2017 had achieved 45.8% coverage, whereas the younger papers (published in 2012) had only achieved 20.7% coverage when sampled in 2021, aged 9. A possible interpretation is that bloggers were moving platforms, possibly to Twitter.

Altmetric started indexing policy documents in 2013, and adds policy repositories on an ongoing basis, which provides an explanation for the apparently growth reported between cohorts, when comparing documents age-for-age. Policy citations are scraped by Altmetric LLP who frequently refresh and grow the policy repositories they access (Altmetric, 2020). In general, this is shown in the data. Nevertheless, policy coverage rates are generally low: papers aged 5 achieve between 1.1% and 5.2% coverage. However, the growth in policy citation across the observation period is very high, with the first policy citations rarely appearing in the first two years of a publication's lifecycle. As policies and white papers are often published at the end of a research and consultative period, this relative lack of pace is to be expected, and is reflected in the much higher coverage figures for the oldest set of publications.

## Growth in coverage over publication age

It has been previously reported that Mendeley saves accumulate from the moment that a paper becomes available (Maflahi and Thelwall, 2018). The observations in this research confirm this finding: sustained growth is observed, Mendeley is a robust, life-long indicator of academic interest. The oldest set of research (published in 2008, sampled in 2021) grew by an average of 5.2 readers in its thirteenth year.

Twitter growth across the age of a paper is less sustained than Mendeley with the rate of coverage expansion dropping off rapidly, confirming observations made elsewhere (Ortega, 2018). Nevertheless, the increase coverage and average values suggest that *both* the number of people tweeting about research *and* the amount of research they share has grown over time.

Two idiosyncrasies are observed. Firstly, Twitter rates for the 2008 cohort increase markedly between years 2014 and 2017, a rise that is not evident in the 2013 cohort. A possible interpretation is that as academic use of Twitter was growing, users were exploring papers published in the preceding decade, whereas papers published in 2013 had already received optimum attention by the Twitter community.

Secondly, there is a small drop in the level of Twitter coverage following the introduction of the General Data Protection Regulation (GDPR) in 2018. This law requires organizations such as Altmetric to remove deleted tweets and Twitter account data. Accordingly, the Twitter numbers following the implementation of GDPR are seen to drop for some cohorts, in terms of both coverage and average.

The rate of news coverage slows with the age of publication but doesn't altogether stop. Although news coverage growth is seen to slow, there are outliers: the oldest set of research acquired new coverage in its thirteenth year. In contrast, blog coverage does plateau, with growth typically slowing by the age 7, and stopping at age 8.

Growth in policy citations is mostly driven by older papers, predominantly in MHS and HSS, suggesting that these two areas have more potential to influence public policy and governance than either LS or PTS. The slow rate of growth suggests two possible interpretations: that the policy process is generally slow and considered, and that research needs to be considered trustworthy - 'tried and tested' - before being incorporated into policy.

COVID-19 has fundamentally changed global health research, both in terms of its speed (Park, Mogg, Smith, Nakimuli-Mpungu et al, 2021) and openness (Fraser, Brierley, Dey, Polka et al, 2021), and it seems likely that policy impact timelines will have been reduced: without longevity of exposure, attention should be given to how research gains trust and authority.

## The probability of late emergence

This research confirms the hypothesis that late-emergent papers may exist (Demain, 2018). Although Mendeley reaches saturation relatively quickly, there are papers getting their first Mendeley readers at ages 5 and 6. Although the likelihood of receiving a first tweet declines dramatically after the first year, this research identified 'first tweets' for papers aged 8 and 9. News and (to a lesser extent) blogs show a more prolonged tail than either Mendeley or Twitter, with the first-year drop being much less marked. Papers without previous mentions, had a 1% chance of receiving their first attention aged 8 and 9.

In contrast with all other indicators, the probability of a paper receiving its first policy citation is seen to rise with the age of the paper: although that probability isn't seen to rise above 1%, this research suggests peak novel policy attention is shown to be between 5-10 years of age.

## The evolution of the OAAA

Mendeley readership reports a consistent OAAA for all years and ages, suggesting that the academic community had a moderate bias towards bookmarking OA research. Twitter coverage (the proportion of papers with tweets) shows no preference between OA and non-OA, however, there is a significant OAAA for the average number of Tweets. Both phenomena suggest that there are two selection processes at play: the active populations are neutral on the question of whether to tweet (or save) an article, but that OA papers are more likely to receive attention once that criterion is met.

The OAAA for news is more complex. In fact, the early years (2008-2010) demonstrate a significant OA Altmetric *Disadvantage*, with parity only being achieved three years into the experiment. Papers published from 2011 did not suffer any disadvantage, and the youngest cohort (published in 2013) enjoyed a significant OAAA ranging from 1.43 to 1.81.

Three elements could be contributing to this effect.

Firstly, as publishing transitions towards increasing rates of OA publications, the quality of OA research could be improving, with commensurate growth in usage.

Secondly, it's possible that the newly emergent OA journals were not investing in preparing press releases, and thereby failing to come to the attention of journalists. The Mendeley data presented here is the best proxy for academic trust, and that suggests that the first explanation - 'lower quality' - is not the answer. Policy – the slowest indicator studied here – shows the weakest and latest OAAA, suggesting that trust in OA research was the slowest to develop.

Thirdly there may be questions of trust or authority involved, with members of the non-academic community unwilling to rely on 'freely available' research, or unfamiliar with the nature or quality of the emergent OA journals.

Whilst care should be taken to communicate the value of all research; the continuing body of evidence for the enhanced and prolonged reach and impact of OA research should taken into account in future research strategy plans.

## Conclusions

This research addresses a number of gaps in the altmetric literature, establishing a number of very long-term trends in platforms: the growth and dominance in Mendeley, and its recent drop in growth; the continued growth of both the research-tweeting population, and the rate at which they tweet, and the reported decline in blogging coverage of research. This research also confirms observations that Policy Citations are the slowest form of attention to accrue, but one that favours the Humanities and Social Sciences

Just as citation databases will add and remove journals from their index and improve their citation parsers, resulting in changes to their data, so altmetric providers make improvements, add new sources and implement legislation. These changes will affect altmetric data: this research may be used to understand those dependencies, and to report on how users may account for the differences.

Late-emergent research is confirmed for the first time for all attention sources, as is the importance of longevity when it comes to the social impact of research via policy documents. Research evaluation techniques should be adopted that properly recognize the slow nature of this valuable impact, especially given the strong bias favouring policy impact for the humanities and social scientists.

This research sheds new light on the adoption and use of OA research, as measured by the existence of different OAAA rates from among the five altmetric attention sources examined. While those sources that are more proximal to the academic community (Mendeley, blogs, Twitter) were relatively quick to show higher usage rates of OA research, those more distant (news, policy) did not show an uptake, and were occasionally seen to be biased against OA research.

These findings reinforce the importance of comparing like-for-like data points, and of normalizing for both year of publication *and* year of collection for altmetric researchers and in metrics calculations; however differences between years do not appear to justify a finer granularity.

## Future work

An area of work hitherto understudied is the degree to which the different attention sources interact with each other, the degree to which these result in broader impact, and the extent to which they are measured by altmetrics. Future work should focus on the mechanisms of social impact, and in particular how implicit assumptions and biases in collection result in uneven attention being paid to certain areas and bodies involved in research. Similarly, the mechanism of interaction between altmetric and citation sources, and how that might differ between Open Access and non-Open Access research has not been analysed in depth.

# References

Allen, H. G., Stanton, T. R., Di Pietro, F., & Moseley, G. L. (2013). Social Media Release Increases Dissemination of Original Articles in the Clinical Pain Sciences. *PLoS ONE*. https://doi.org/10.1371/journal.pone.0068914

Almind, T. C., & Ingwersen, P. (1997). Informetric analyses on the world wide web: methodological approaches to 'webometrics.' *Journal of Documentation*, *53*(4), 404–426. https://doi.org/10.1108/EUM0000000007205

Altmetric. (2018). Patent data in Altmetric highlights the commercialization of research – Altmetric. Retrieved December 16, 2019, from https://www.altmetric.com/press/press-releases/patent-data-in-altmetric-highlights-the-commercialization-of-research/

Altmetric. (2020). FAQ Policy Documents. Retrieved July 29, 2021, from https://help.altmetric.com/support/solutions/articles/6000236695-policy-documents

Bar-Ilan, J. (2000). The web as an information source on informetrics? A content analysis. *Journal of the American Society for Information Science and Technology*, *51*(5), 432–443. https://doi.org/10.1002/(sici)1097-4571(2000)51:5<432::aid-asi4>3.0.co;2-7

Björk, B.-C., Welling, P., Laakso, M., Majlender, P., Hedlund, T., & Guðnason, G. (2010). Open Access to the Scientific Journal Literature: Situation 2009. *PLoS ONE*, *5*(6), e11273. https://doi.org/10.1371/journal.pone.0011273

Braun, T., Glänzel, W., & Schubert, A. (2010). On Sleeping Beauties, Princes and other tales of citation distributions …. *Research Evaluation* 19(3). https://doi.org/10.3152/095820210x514210

Clements, A., Darroch, P. I., & Green, J. (2017). Snowball Metrics – Providing a Robust Methodology to Inform Research Strategy – but do they help? *Procedia Computer Science*, *106*, 11–18. https://doi.org/10.1016/J.PROCS.2017.03.003

Demaine, J. (2018). Rediscovering Forgotten Research: Sleeping Beauties at the University of Waterloo. *J Inf Sci Theory Pract*, 6(3) 37–44. https://doi.org/10.1633/JISTaP.2018.6.3.4

Deville, S. (2013) To Mendelete or Not to Mendelete? Retrieved July 29, 2021, from https://sylvaindeville.net/2013/04/10/to-mendelete-or-not-to-mendelete/

Ebrahimy, S., Mehrad, J., Setareh, F., & Hosseinchari, M. (2016). Path analysis of the relationship between visibility and citation: the mediating roles of save, discussion, and recommendation metrics. *Scientometrics*, *109*(3), 1497–1510. https://doi.org/10.1007/s11192-016-2130-z

Eng, D. (2017) Toronto at the Centre of the Race to Save Climate Change Data from Trump. *Torontoist*. Accessed July 30, 2021 https://torontoist.com/2017/04/toronto-centre-race-save-climate-change-data-trump/

European Union (2016) *Regulation (EU) 2016/679 of the European Parliament and of the Council* https://eur-lex.europa.eu/eli/reg/2016/679/oj

Eysenbach, G. (2011). Can tweets predict citations? Metrics of social impact based on Twitter and correlation with traditional metrics of scientific impact. *Journal of Medical Internet Research.* 13(4) https://doi.org/10.2196/jmir.2012

Fang, Z,. Costas, R. (2020). Studying the accumulation velocity of altmetric data tracked by Altmetric.com. *Scientometrics,* 123. https://doi.org/10.1007/s11192-020-03405-9

Fang, Z., Costas, R., Tian, W., Wang, X., & Wouters, P. (2020). An extensive analysis of the presence of altmetric data for Web of Science publications across subject fields and research topics. Scientometrics, 124(3), 2519–2549. https://doi.org/10.1007/S11192-020-03564-9

Fraser, N., Brierley, L., Dey, G., Polka, J.K., Pálfy, M., et al. (2021) The evolving role of preprints in the dissemination of COVID-19 research and their impact on the science communication landscape. PLOS Biology 19(4): e3000959. https://doi.org/10.1371/journal.pbio.3000959

Fraumann, G., & Colavizza, G. (2022). The role of blogs and news sites in science communication during the COVID-19 pandemic. Frontiers in Research Metrics and Analytics, 7. https://doi.org/10.3389/FRMA.2022.824538

Gorraiz, J., Blahous, B., & Wieland, M. (2018). Monitoring the broader impact of the journal publication output on country level: A case study for Austria. *Communications in Computer and Information Science*, *856*, 39–62. https://doi.org/10.1007/978-981-13-1053-9_4/TABLES/13

Hawkins, C. M., Hillman, B. J., Carlos, R. C., Rawson, J. V., Haines, R., & Duszak, R. (2014). The impact of social media on readership of a peer-reviewed medical journal. *Journal of the American College of Radiology* 11(11). https://doi.org/10.1016/j.jacr.2014.07.029

Haustein, S. (2016). Grand challenges in altmetrics: heterogeneity, data quality and dependencies. *Scientometrics,* 108. https://doi.org/10.1007/s11192-016-1910-9

Herzog, C., Sorensen, A., & Taylor, M. (2016). Forward-looking analysis based on grants data and machine learning based research classifications as an analytical tool, 1–10. Retrieved from https://www.oecd.org/sti/093 - OECDForward-lookinganalysisbasedongrantsdataandmachinelearningbasedresearchclassificationsasananalyticaltool (1).pdf

Holmberg, K., Hedman, J., Bowman, T. D., Didegah, F., & Laakso, M. (2020). Do articles in open access journals have more frequent altmetric activity than articles in subscription-based journals? An investigation of the research output of Finnish universities. *Scientometrics*, 122. https://doi.org/10.1007/s11192-019-03301-x

Hou, J., Li, H., & Zhang, Y. (2020). Identifying the princes base on Altmetrics: An awakening mechanism of sleeping beauties from the perspective of social media. PLoS ONE, 15(11 November). https://doi.org/10.1371/JOURNAL.PONE.0241772

Htoo, T. H. H., Jin-Cheon, N., & Thelwall, M. (2022). Why are medical research articles tweeted? The news value perspective. Scientometrics. https://doi.org/10.1007/S11192-022-04578-1

Hutchins, B. I., Yuan, X., Anderson, J. M., & Santangelo, G. M. (2016). Relative Citation Ratio (RCR): A New Metric That Uses Citation Rates to Measure Influence at the Article Level. *PLoS Biology*, *14*(9). https://doi.org/10.1371/JOURNAL.PBIO.1002541

Jamali, H. R. & Alimohammadi, D. (2015). Blog Citations as Indicators of the Societal Impact of Research: Content Analysis of Social Sciences Blogs. *International Journal of Knowledge Content Development & Technology*, *5*(1), 15–32. https://doi.org/10.5865/ijkct.2015.5.1.015

Kudlow, P., Cockerill, M., Toccalino, D., Dziadyk, D. B., Rutledge, A., Shachak, A., … Eysenbach, G. (2017). Online distribution channel increases article usage on Mendeley: a randomized controlled trial. *Scientometrics*, *112*(3), 1537–1556. https://doi.org/10.1007/s11192-017-2438-3

Lee, J. L. & Haupt, J. P. (2020) Scientific globalism during a global crisis: research collaboration and open access publications on COVID-19, *Higher Education*. https://doi.org/10.1007/s10734-020-00589-0

Maflahi, N., & Thelwall, M. (2018). How quickly do publications get read? The evolution of mendeley reader counts for new articles. *Journal of the Association for Information Science and Technology*. https://doi.org/10.1002/asi.23909

McLeish, B. (Altmetric). (2016, September). Altmetric and Policy: Discovering how your research impacted real-world practises. Retrieved December 16, 2019, from https://www.altmetric.com/blog/altmetric-and-policy-discovering-how-your-research-impacted-real-world-practises/

Moed, H. F. (2005). Statistical relationships between downloads and citations at the level of individual documents within a single journal. *Journal of the American Society for Information Science and Technology*, 56(10). https://doi.org/10.1002/asi.20200

Mohammadi, E., Thelwall, M. & Kousha, K. (2016). Can Mendeley bookmarks reflect readership? A survey of user motivations. *Journal of the Association for Information Science and Technology*, 67(5), 1198-1209. doi:10.1002/asi.23477

Ortega, J. L. (2018). The life cycle of altmetric impact: A longitudinal study of six metrics from PlumX. *Journal of Informetrics*, *12*, 579–589. https://doi.org/10.1016/j.joi.2018.06.001

Park, J. J. H., Mogg, R., Smith, G. E., Nakimuli-Mpungu, E., Jehan, F. Rayner, C. R. Condo, J., Decloedt, E. H., Nachega, J. B., Reis, G., Mills, E. J. (2021). How COVID-19 has fundamentally changed clinical research in global health. The Lancet Global Health https://doi.org/10.1016/S2214-109X(20)30542-8

Perneger, T. V. (2004). Relation between online "hit counts" and subsequent citations: Prospective study of research papers in the BMJ. *BMJ*. https://doi.org/10.1136/bmj.329.7465.546

Phillips, D. P., Kanter, E. J., Bednarczyk, B., & Tastard, P. L. (1991). Importance of the Lay Press in the Transmission of Medical Knowledge to the Scientific Community. The New England Journal of Medicine https://doi.org/10.1056/NEJM199110173251620

Piwowar, H., Priem, J., Larivière, V., Alperin, J. P., Matthias, L., Norlander, B., … Haustein, S. (2018). The state of OA: a large-scale analysis of the prevalence and impact of Open Access articles. *PeerJ*, *6*, e4375. https://doi.org/10.7717/peerj.4375

Priem, J., Taraborelli, D., Groth, P., & Neylon, C. (2010). Alt-metrics: a manifesto. Retrieved from http://altmetrics.org/manifesto/

Schlögl, C., Gorraiz, J., Gumpenberger, C., Jack, K., & Kraker, P. (2014). Comparison of downloads, citations and readership data for two information systems journals. *Scientometrics*, 101. https://doi.org/10.1007/s11192-014-1365-9

Shema, H., Bar-Ilan, J., & Thelwall, M. (2012). Research blogs and the discussion of scholarly information. PLoS ONE, 7(5), e35869. https://doi.org/10.1371/journal.pone.0035869

Shuai, X., Pepe, A., & Bollen, J. (2012). How the Scientific Community Reacts to Newly Submitted Preprints: Article Downloads, Twitter Mentions, and Citations. *PLoS ONE*. https://doi.org/10.1371/journal.pone.0047523

Statista. (2021) Annual Twitter user growth rate worldwide from 2015 to 2013. Retrieved July 29, 2021, from https://www.statista.com/statistics/303723/twitters-annual-growth-rate-worldwide/

Suber, P. (2012). *Open access*, MIT Press. Retrieved from https://mitpress.mit.edu/books/open-access

Sugimoto, C. R., Work, S., Larivière, V., Haustein, S. (2017). Scholarly Use of Social Media and Altmetrics: A Review of the Literature. Journal of the Association for Information Science and Technology, 68 (9), 2027-2062. https://doi.org/10.1002/asi.23833

Taylor, M. (2014). Open methods: bringing transparency to research metrics, Septentrio Conference Series. https://doi.org/10.7557/5.3240

Taylor, M. (2020). An Altmetric Attention Advantage for Open Access Books in the Humanities and Social

Sciences, *Scientometrics*. https://doi.org/10.1007/s11192-020-03735-8

Thelwall, M. (2000). Web impact factors and search engine coverage. *Journal of Documentation*, *56*(2), 185–189. https://doi.org/10.1108/00220410010803801

Thelwall, M. (2016). Interpreting correlations between citation counts and other indicators. *Scientometrics*, *108*(1). https://doi.org/10.1007/s11192-016-1973-7

Thelwall, M. (2017). Three practical field normalised alternative indicator formulae for research evaluation. *Journal of Informetrics*, *11*(1), 128–151. https://doi.org/10.1016/j.joi.2016.12.002

Thelwall, M., & Fairclough, R. (2015). Geometric journal impact factors correcting for individual highly cited articles. *Journal of Informetrics*. https://doi.org/10.1016/j.joi.2015.02.004

Thelwall, M., Haustein, S., Larivière, V., & Sugimoto, C. R. (2013). Do altmetrics work? Twitter and ten other social web services. *PloS One*, *8*(5), e64841. https://doi.org/10.1371/journal.pone.0064841

Thelwall, M., & Sud, P. (2016). Mendeley readership counts: An investigation of temporal and disciplinary differences. *Journal of the Association for Information Science and Technology*. https://doi.org/10.1002/asi.23559

Watson, A. B. (2009). Comparing citations and downloads for individual articles. *Journal of Vision*, *9*(4), 1–4. https://doi.org/10.1167/9.4.i

Wheatley, D. & Grzynszpan, D. (2002). Can we speed up the online publishing process? And who will pay for it, anyway? *Cancer Cell International*, *2*(1), 5. https://doi.org/10.1186/1475-2867-2-5

Williams, K. (2018). Three strategies for attaining legitimacy in policy knowledge: Coherence in identity, process and outcome. *Public Administration*, *96*(1), 53–69. https://doi.org/10.1111/padm.12385

Zhang, G., Wang., Y., Weixi, X., Du, H., Jiang, C., Wang. (2020) The open access usage advantage: a temporal and spatial analysis. *Scientometrics* https://doi.org/10.1007/s11192-020-03836-4

Zahedi, Z., Costas, R., Wouters, P. (2014). How well developed are altmetrics? A cross-disciplinary analysis of the presence of 'alternative metrics' in scientific publications. *Scientometrics.* https://doi.org/10.1007/s11192-014-1264-0

Zanotto, E. D., & Carvalho, V. (2021). Article age- and field-normalized tools to evaluate scientific impact and momentum. *Scientometrics*, *126*(4), 2865–2883. https://doi.org/10.1007/S11192-021-03877-3

# Appendix

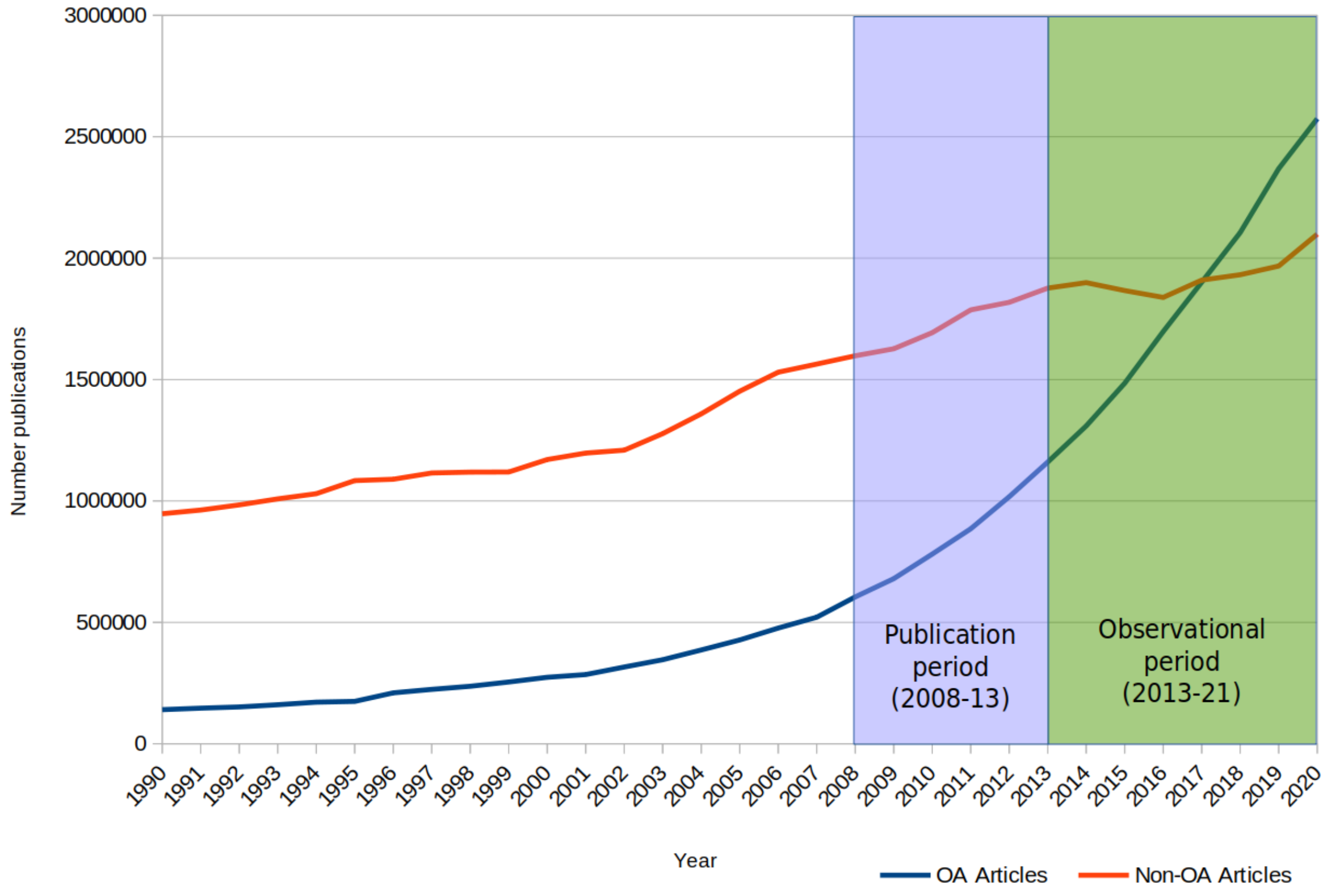

***Fig. 12*** *The trends in OA publishing rate over time*